\documentclass[twocolumn]{aastex631}
\usepackage{amsmath}

\begin{document}

\title{Reverse Shock Emission from Misaligned Structured Jets in Gamma-Ray Bursts}

\author[0009-0002-0137-2650]{Sen-Lin Pang}
\affiliation{Department of Astronomy, University of Science and Technology of China, Hefei 230026, China; daizg@ustc.edu.cn}
\author[0000-0002-7835-8585]{Zi-Gao Dai}
\affiliation{Department of Astronomy, University of Science and Technology of China, Hefei 230026, China; daizg@ustc.edu.cn}
\affiliation{School of Astronomy and Space Science, University of Science and Technology of China, Hefei 230026, China}

\begin{abstract}
The afterglow of gamma-ray bursts (GRBs) has been extensively discussed in the context of shocks generated during an interaction of relativistic outflows with their ambient medium. This process leads to the formation of both a forward and a reverse shock. While the emission from the forward shock, observed off-axis, has been well-studied as a potential electromagnetic counterpart to a gravitational wave-detected merger, the contribution of the reverse shock is commonly overlooked. In this paper, we investigate the contribution of the reverse shock to the GRB afterglows observed off-axis. In our analysis, we consider jets with different angular profiles, including two-component jets, power-law structured jets, Gaussian jets and 'mixed jets' featuring a Poynting-flux-dominated core surrounded by a baryonic wing. We apply our model to GRB 170817A/GW170817 and employ the Markov Chain Monte Carlo (MCMC) method to obtain model parameters. Our findings suggest that the reverse shock emission can significantly contribute to the early afterglow. In addition, our calculations indicate that the light curves observable in future off-axis GRBs may exhibit either double peaks or a single peak with a prominent feature, depending on the jet structure, viewing angle and micro-physics shock parameters.
\end{abstract}


\section{Introduction} \label{sec:intro}
Gamma-ray bursts (GRBs) stand as the most energetic explosions in the universe,
typically arising from the energy dissipation within ultra-relativistic jets launched during the core-collapse of massive stars \citep{Wang_1998,Woosley&Bloom2006,cano2017} or from the merger of two compact objects like two neutron stars (NS-NS) \citep{Paczynski_1986,Thompson1994}, a neutron star and a stellar-mass black hole (NS-BH) \citep{Narayan1992}, or a neutron star and a white dwarf \citep{Yang2022,Zhong_2023}.
GRB observations typically involve two phases: prompt emission and afterglow emission.
The prompt gamma-ray emission of a GRB can persist for duration ranging from a fraction of one second to several minutes, while the afterglow can be observed across multiple wavelengths for months, spanning from X-rays to radio frequencies, and even extending into the realm of very-high-energy gamma rays.
Although the exact physics governing the prompt emission remains uncertain (e.g., internal shock or ICMART) \citep{internal_shock, Zhang_2011_ICMART}, a generalized synchrotron external shock model has been established to interpret broad-band afterglow observations.

The external shock model delineates an interaction between the relativistic GRB jet and its circum-bust medium (CBM).
During this interaction, two shocks emerge: a forward shock (FS) and a reverse shock (RS), sweeping into the CBM and the jet itself, respectively \citep{Sari_1995,Meszaros_1999,Yi_2013,Yi_2020}. 
Electrons undergo acceleration in both shocks and release their energy in the magnetic fields generated by plasma instabilities \citep{Medvedev_1999}. 
Once the RS crosses the jet shell, the shocked CBM and the shocked ejecta can be treated as a blastwave region, and their evolution is solved using the Blandford-McKee self-similar solutions \citep{BM76}.
In this paper, we employ the semi-analytical solution proposed by \citet{Zhang_Ze_lin_2022} to model the evolution of the FS-RS system. Once the RS crosses the shell, we utilize a generic dynamical model introduced by \citet{Huang_1999} to describe the evolution of the FS.

Studies on off-axis GRBs commenced early.
\citet{Woods_1999} calculated the prompt X-ray emission as a function of the view angle for beamed GRB sources.
\citet{Huang_2004} use an off-axis two-component jet to explain the rebrightening of XRF 030723.
Subsequently, \citet{Granot_2005} focused on calculating the FS emission from structured jets in off-axis GRBs.
Significant attention was garnered by the observations of GRB 170817A/GW170817, posited as the first GRB clearly viewed far from the jet's symmetry axis \citep{Abbott_2019,Makhathini_2021}. This event has predominantly been explained through FS emission from an off-axis structured jet, although the contribution of the reverse shock (RS) has not been factored in \citep{Lazzati2017,Gill_2018}.
However, \citet{Fraija_2019} conducted a study on the off-axis emission of GRB 170817A, demonstrating that the observed $\gamma$-ray flux could be consistent with a synchrotron self-Compton (SSC) model involving the RS.
Additionally, \citet{Lamb_2019} calculated the RS emission from an off-axis viewpoint, suggesting that the distinctive feature in the pre-peak afterglow could be attributed to the RS.
Furthermore, \citet{lamb2020reverse} use the RS emission to explain the radio afterglow of slightly off-axis GRB 160821B.

In our previous work \citep{Pang_2024}, we developed a top-hat jet forward-reverse shock emission model, and fitted the afterglows of  GRB 080503, GRB 140903A, GRB 150101B, GRB 160821B, and GW 170817/GRB 170817A.
In this paper, we investigate the emission of the FS-RS system from misaligned structured jets associated with GRBs in both ISM and wind environments. We then apply our model to fit the observations of confirmed off-axis GRB 170817A. 
Comparing to the work of \citet{Lamb_2019}, we use more data from GRB 170817A and find that the RS may contribute to the early radio and X-ray afterglow. 
Furthermore, we compare the light curves for thick shell and thin shell cases by discussing how the thickness of the ejecta shell influence the light curves.

This paper is organized as follows.
In Section~\ref{subsec:DR}, we provide an overview of the dynamics and radiation process of relativistic GRB jets. 
In Section~\ref{subsec:LC}, we present the calculations of the emission from an misaligned structured jet and show the examples of afterglow light curves. 
And in Section~\ref{subsec:Mix_jet}, we calculate the emission from "mixed jet" consisting of a Poynting-flux-dominated core surrounded by a baryonic wing.
In Section~\ref{sec:results}, we apply our model to fit the data from confirmed off-axis GRB 170817A and present the fitting results. 
Finally, Section~\ref{sec:4} contains our discussion and conclusions based on findings of this study.

\section{The Model} \label{sec:model}
\subsection{Jet dynamics and radiation processes} \label{subsec:DR}
We consider a scenario where a cold shell, characterized by an isotropic-equivalent energy $E_{\rm iso}$ and an initial Lorentz factor $\Gamma_0$ ejected from a central engine, expands in a cold circum-burst medium (CBM).
The number density of the CBM is parameterized as $n_1(R)=AR^{-k}$, where $k = 0$ and $2$ correspond to the ISM and wind environments, respectively. 
In the case of an ISM, $A=n_0$, which is typically set to $n_0 \sim 1 \mathrm{cm}^{-3}$.
For a wind environment, $A$ can be expressed as \citep{Chevalier&Li_1999}
\begin{equation}
    A = \frac{\dot{M}}{4\pi m_\mathrm{p} v_\mathrm{w}} = 3\times 10^{35} A_* \mathrm{~cm}^{-1},
    \label{eq_A}
\end{equation}
where $\dot{M}$ is the mass-loss rate of the progenitor star, typically $\dot{M}\sim10^{-5}M_\odot/\mathrm{yr}$, and $v_\mathrm{w}$ is the velocity of the wind around $v_\mathrm{w}\sim 10^3 \mathrm{km}/\mathrm{s}$.
The dimensionless parameter $A_*$ is defined as $A_*=(\dot{M}/10^{-5}M_\odot\mathrm{yr}^{-1})(v_\mathrm{w}/10^3\mathrm{km}\mathrm{~s}^{-1})$.

When a relativistic shell interacts with its CBM, an FS-RS system are formed. We divide the system into four regions \citep{Sari_1995}:
(1) the unshocked CBM ($\Gamma_1=0$,$e_1=0$, $n_1$), (2) the forward-shocked CBM ($\Gamma_2$, $e_2'$, $n_2'$), (3) the reverse-shocked shell ($\Gamma_3$, $e_3'$, $n_3'$), and (4) the unshocked shell ($\Gamma_4=\Gamma_0$, $n_4'$), where $e'$ and $n'$ are internal energy density and number density in co-moving frame.
We assume uniform bulk Lorentz factors for the shocked regions, denoted as $\Gamma_2=\Gamma_3=\Gamma$.
Throughout this paper, the single script index $i$ denotes the quantities in Region $i$, and the double index $ij$ denotes the relative value between Region $i$ and $j$.
Additionally, the superscript prime ($'$) indicates the quantities in the co-moving frame.
Two limiting cases are commonly discussed: the thick-shell case corresponds to a relativistic reverse shock (RRS), while the thin-shell case corresponds to a non-relativistic (Newtonian) reverse shock (NRS).
In the generic case, corrections to the maximum flux, synchrotron peak, and cooling frequencies have been proposed by \citet{Nakar&Piran2004} and \citet{Harrison_2013}. 
In this paper, we use the semi-analytical solutions developed by \citet{Zhang_Ze_lin_2022} to address the generic case.

Following \citet{Sari_1995}, we introduce dimensionless quantity,
\begin{equation}
    \xi \equiv (3-k)^{-\frac{1}{2(3-k)}} \left(\frac{l}{\Delta_0}\right)^{1/2}\Gamma_0^{-\frac{4-k}{3-k}},
    \label{eq_xi}
\end{equation}
where $\Delta_0$ represents the initial shell width in the lab frame, $l=\left[(3-k)E_\mathrm{iso}/(4\pi Am_pc^2)\right]^{1/(3-k)}$ is the Sedov length. The condition $\xi\ll1$ characterizes the thick-shell case, while $\xi\gg1$ describes the thin-shell case. 
According to energy and conservation and shock jump conditions, with $\Gamma\gg 1$, we have \citep{Pe'er_2012,Zhang_Ze_lin_2022,Pang_2024}
\begin{equation}
\begin{aligned}
    \Gamma = \Gamma_0 \sqrt{\frac{m_3c^2}{2\Gamma_0 m_2 c^2 + m_3c^2}}
\end{aligned}
\label{eq_dyn1}
\end{equation}
where $m_2 = 4\pi A m_\mathrm{p} R^{3-k}/(3-k)$ represents the mass swept by the FS, $m_\mathbf{ej}=E_{\rm iso}/(\Gamma_0c^2)$ is the initial mass of the shell. 
The mass of the shocked shell, $m_3$, varies depending on whether the system is in the thick-shell or thin-shell case.
This difference in $m_3$ expressions leads to the distinct evolution of $\Gamma$ in these two limiting cases.
However, \citet{Zhang_Ze_lin_2022} provided a general solution for the Lorentz factor $\Gamma$ during the RS crossing phase with reasonable accuracy, based on the condition of energy conservation. 
According to the semi-analytical expressions from \citet{Zhang_Ze_lin_2022} (Table 2), the RS crossing radius can be approximate as
\begin{equation}
    \frac{R_\Delta}{\Gamma_0^2 \Delta_0} \approx (f_1^s + f_2^s)^{1/s},
    \label{eq_R_d}
\end{equation}
where $f_1 = \left[ \frac{(3-k)(4-k)}{4}\xi^{2(3-k)} \right]^{1/(4-k)}$ corresponds to the thick-shell case,  while $f_2=\left[ \frac{3(3-k)^2}{8}\xi^{2(3-k} \right]^{1/(3-k)}$ corresponds the thin-shell case. 
Denoting $x\equiv R/R_\Delta$, before the RS crosses the ejecta (i.e. $x<1$), the evolution of Lorentz factor can be approximate as 
\begin{equation}
    \begin{aligned}
        \Gamma = \Gamma_0&\left[ 1 + \left(\frac{4-k}{2}\right)^{\frac{2-k}{4-k}}\left(\frac{4-k}{3-k}\right)^{\frac{1}{4-k}}\xi^{-\frac{2(3-k)}{4-k}}x^{\frac{2-k}{2}}\right. \\
        &+ \left. \frac{3(3-k)}{4}x^{\frac{(3-k)}{2}} \right]^{-1/2}.
    \end{aligned}
    \label{eq_Gamma1}
\end{equation}
For $\xi \ll 1$ and $\xi \gg 1$, this equation simplifies to describe the thick-shell and thin-shell cases, respectively.

Once the RS crosses the shell (i.e., when $m_3=m_\mathrm{ej}$), according to energy conservation and shock jump conditions under the adiabatic approximation, the Lorentz factor of FS $\Gamma_2$ can be expressed as  \citep{Huang_1999,Pang_2024}
\begin{equation}
    \Gamma_2^2m_2 + \Gamma_2\Gamma_{34,\Delta}m_{\mathrm{ej}} - (\Gamma_\Delta^2\Gamma_{34,\Delta}m_\mathrm{ej} + m_2 - m_\mathrm{ej}) = 0,
    \label{eq_dyn2}
\end{equation}
where the subscript $\Delta$ denotes the quantities at crossing radius $R_\Delta$.

After RS crosses the shell, the shocked shell can be approximately treated as the tail of FS. Following the Blandford-McKee (BM) solution, the evolution of the shocked shell proceeds as $\Gamma_3 \propto R^{(2k-7)/2}$, $e_3'\propto R^{(4k-26)/3}$ and $n_3'\propto R^{(2k-13)/2}$ \citep{Kobayashi&Sari_2000, Yi_2013}.
However, the BM solutions may exhibit deviations because the temperature of the shocked ejecta may not be relativistic for a Newtonian or mildly relativistic RS. According to the numerical study by \citet{Kobayashi&Sari_2000}, the two extreme cases of cold and hot shells yield similar light curves, which has no significant impact on our results.

Assuming that the distributions of electrons accelerated by the FS and RS follow power laws of an index $p$ (i.e., $N_\mathrm{e}(\gamma_\mathrm{e})\propto \gamma_\mathrm{e}^{-p}$), with the minimum Lorentz factor of the shocked electrons in the shell rest frame for $p>2$ is given by \citep{SPN98,Huang_2000}
\begin{equation}
    \gamma_{\mathrm{m},i} = \epsilon_{\mathrm{e},i}\frac{m_\mathrm{p}}{m_\mathrm{e}}\frac{p-2}{p-1}(\Gamma_\mathrm{sh}-1) + 1,
    \label{eq_gamma_e_min}
\end{equation}
where $\epsilon_{\mathrm{e},i}$ is the fraction of the internal energy in Region $i$ carried by electrons, $\Gamma_\mathrm{sh}$ is the relative upstream to downstream Lorentz across the shock ($\Gamma_\mathrm{sh}=\Gamma_2$ for the forward-shock region, and $\Gamma_\mathrm{sh}=\Gamma_{34}$ for the reverse-shock region). 
When the shock become Newtonian then $\Gamma_\mathrm{sh}-1\approx \beta_\mathrm{sh}^2/2$, and
\begin{equation}
    \gamma_{\mathrm{m},i}\approx \epsilon_{\mathrm{e},i} \frac{m_\mathrm{p}}{m_\mathrm{e}} \frac{p-2}{p-1} \frac{\beta_\mathrm{sh}^2}{2} + 1.
    \label{eq_gamma_e_m_NR}
\end{equation}
For $1<p<2$, the maximum electron Lorentz factor is calculated by assuming that the acceleration time equals the synchrotron time, it is given by \citep{Dai_&_Cheng_2001}
\begin{equation}
    \gamma_{M,i}=\sqrt{\frac{3q_e}{\zeta\sigma_TB_i'}},
    \label{eq_gamma_max}
\end{equation}
where $q_e$ is the electron charge, $\sigma_T$ denotes the Thomson cross-section, $\zeta\sim 1$ is the ratio of the acceleration time to the gyration time, and $B_i'$ is magnetic strength.
Assuming that $\epsilon_{B,i}$ is the fraction of the internal energy in Region $i$ carried by the magnetic field, the magnetic energy density in the jet’s co-moving frame can thus be estimated as $B'^2/(8\pi) = \epsilon_{B,i}e_i'$ \citep{SPN98}. 
Assuming that $\gamma_M\gg \gamma_\mathrm{m}$, the minimum Lorentz factor of shocked electrons for $1<p<2$ is given by
\begin{equation}
    \gamma_{\mathrm{m},i}=\left[\epsilon_{\mathrm{e},i} \frac{2-p}{p-1} \frac{m_\mathrm{p}}{m_\mathrm{e}} \gamma_{M,i}^{p-2} (\Gamma_\mathrm{sh}-1)\right]^{1/(p-1)} + 1.
\end{equation}
Additionally, the Lorentz factor of the electrons that cool on a timescale equal to the dynamic timescale is given by \citep{SPN98}
\begin{equation}
    \gamma_\mathrm{c} = \frac{6\pi m_\mathrm{e}c}{(Y+1)\sigma_T\Gamma B'^2 t}.
\end{equation}
In the above equation, $Y$ denotes the Compton parameter, representing the ratio of the inverse-Compton (IC) luminosity to the synchrotron luminosity, and can be calculated by $Y_i = (-1+\sqrt{1+4\xi_e\epsilon_{e,i}\epsilon_{B,i}})/2$ with $\xi_e$ being the fraction of the electron energy radiated.

The local co-moving synchrotron emissivity $P_{\nu'}'$ can be expressed as a broken power law \citep{SPN98,Gao_2013}. 
In the slow-cooling regime, 
\begin{equation}
    \frac{P_{\nu'}'}{P_{\nu',\mathrm{max}}'}=\begin{cases}
        (\nu_\mathrm{a}'/\nu_\mathrm{m}')^{1/3}(\nu'/\nu_\mathrm{a}')^2, &\nu'<\nu_\mathrm{a}'<\nu_\mathrm{m}'\\
        (\nu_\mathrm{m}'/\nu_\mathrm{a}')^{(p+4)/2}(\nu'/\nu_\mathrm{m}')^2, &\nu'<\nu_\mathrm{m}'<\nu_\mathrm{a}'\\
        (\nu'/\nu_\mathrm{m}')^{1/3}, &\nu_\mathrm{a}'<\nu'<\nu_\mathrm{m}'\\
        (\nu_\mathrm{a}'/\nu_\mathrm{m}')^{-(p-1)/2}(\nu'/\nu_\mathrm{a}')^{5/2}, &\nu_\mathrm{m}'<\nu'<\nu_\mathrm{a}'\\
        (\nu'/\nu_\mathrm{m}')^{-(p-1)/2}, &\nu_\mathrm{a}',\nu_\mathrm{m}'<\nu'<\nu_\mathrm{c}'\\
        (\nu_\mathrm{c}'/\nu_\mathrm{m}')^{-(p-1)/2}(\nu'/\nu_\mathrm{c}')^{-p/2}, &\nu_\mathrm{c}'<\nu'
    \end{cases}
        \label{eq_syn1}
\end{equation}
and in the fast-cooling regime,
\begin{equation}
    \frac{P_{\nu'}'}{P_{\nu',\mathrm{max}}'}=\begin{cases}
        (\nu_\mathrm{a}'/\nu_\mathrm{c}')^{1/3}(\nu'/\nu_\mathrm{a}')^2, &\nu'<\nu_\mathrm{a}'\\
        (\nu'/\nu_\mathrm{c}')^{1/3}, &\nu_\mathrm{a}'<\nu'<\nu_\mathrm{c}'\\
        (\nu'/\nu_\mathrm{c}')^{-1/2}, &\nu_\mathrm{c}'<\nu'<\nu_\mathrm{m}'\\
        (\nu_\mathrm{m}'/\nu_\mathrm{c}')^{-1/2}(\nu'/\nu_{\mathrm{m}}')^{-p/2}, &\nu_\mathrm{m}'<\nu',
    \end{cases}
        \label{eq_syn2}
\end{equation}
where the first characteristic frequency $\nu_\mathrm{a}'$ is the self-absorption frequency calculated as $k_{\nu_\mathrm{a}'}L'=1$ \citep{Wu_2003}. Here, $k_{\nu_\mathrm{a}'}$ is the self-absorption coefficient at $\nu_\mathrm{a}'$, and $L'$ is the length of the radiation region in the co-moving frame.
$\nu_\mathrm{m}'$ and $\nu_\mathrm{c}'$ are characteristic frequencies corresponding to the co-moving frame minimum electron Lorentz factor $\gamma_\mathrm{m}$, and cooling Lorentz factor $\gamma_\mathrm{c}$, respectively.
The flux normalization and break frequencies are \citep{SPN98,Wijers&Galama_1999}
\begin{subequations}
    \begin{align}
        P_{\nu',\mathrm{max},i}'&\simeq \frac{m_e c^2\sigma_T}{3q_e}(8\pi)^{1/2}\epsilon_{B,i}^{1/2}e_i'^{1/2},
            \label{eq_P}\\
        \nu_{\mathrm{m},i}'&=\frac{3x_p}{4\pi}\gamma_{\mathrm{m},i}^2\frac{q_eB_i'}{m_ec},
            \label{eq_nu_m}\\
        \nu_{\mathrm{c},i}'&= 0.286\frac{3}{4\pi}\gamma_{\mathrm{c},i}^2 \frac{q_eB_i'}{m_ec}
            \label{eq_nu_c},
    \end{align}
\end{subequations}
where $x_p$ is a dimensionless factor that depends on the electron power-law $p$.

In addition to synchrotron radiation, we also account for inverse-Compton (IC) scattering radiation from shocked electrons.
The IC radiation comprises two parts: synchrotron self-Compton (SSC) radiation and combined-IC radiation \citep{Wang_2001}, where photons from Region $j$ can be scattered by electrons in Region $i$ (with $j=i$ for SSC and $j\ne i$ for combined-IC). 
The IC volume emissivity in the rest frame of shell can be calculated as \citep{Wang_2001} 
\begin{equation}
    j^{\prime \mathrm{IC}}_{\nu',i} = 3\sigma_T \int_{\gamma_{\mathrm{min},i}'}^{\gamma_{\mathrm{max},i}'}\mathrm{d}\gamma_{e,i}'N_{e,i}(\gamma_{e,i}')\int_0^1\mathrm{d}xg(x)\bar{f}_{\nu'_s,j}'(x),
        \label{eq_IC}
\end{equation}
where $x\equiv \nu'/(4\gamma_{e,i}'^2\nu_{s,j}')$, $\bar{f}'_{\nu'_s}$ represents the incident-specific synchrotron flux at shock front, and $g(x)=1+x+2x\ln(x)-2x^2$.

\subsection{Light curves from a misaligned structured jet} \label{subsec:LC}
\begin{figure}
    \centering
    \includegraphics[width=\columnwidth]{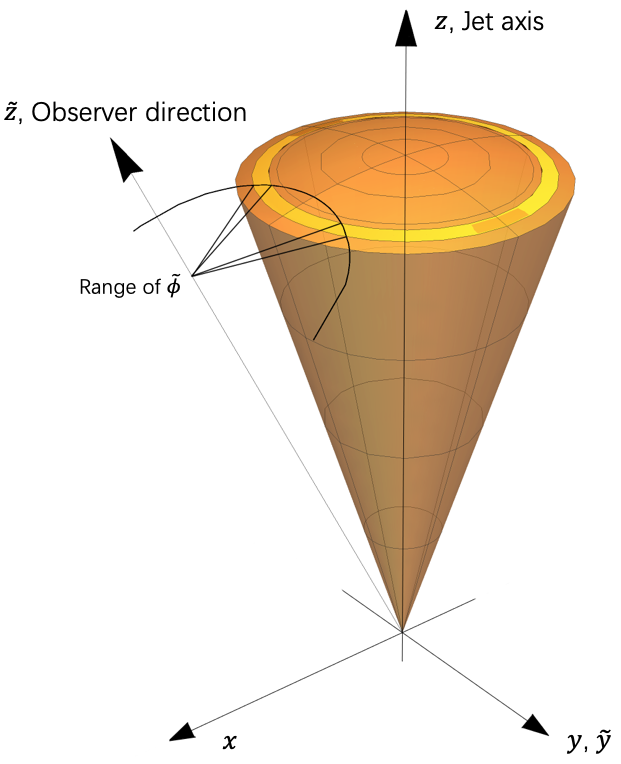}
    \caption{The coordinate system used to calculate afterglow emission from the observer direction. The jet direction is along the $z$ direction, while LOS is along the $\Tilde{z}$ direction.
    The yellow ring is an arbitrary segment of structured jet.}
    \label{fig1}
\end{figure}
Consider an infinitely thin-shell and neglect the radial structure of emitting regions. The emission originates within polar angles $0\le \theta\le\theta_\mathrm{j} $, where $\theta_\mathrm{j}$ represents the jet's half-opening angle. We conveniently align the jet's symmetry axis with the $z$ direction, and the direction towards the observer, $\hat{\Tilde{z}}$, forms an angle $\theta_\mathrm{obs} = \arccos(\hat{\Tilde{z}}\cdot\hat{z})$ relative to the jet axis (see Fig.~\ref{fig1}).
For a given observed time $t_\mathrm{obs}$, the afterglow emission is derived by integrating the emissivity over the equal-arrival-time-surface (EATS), which is determined by using the relation \citep{Gill_2018}
\begin{equation}
    t_z\equiv\frac{t_\mathrm{obs}}{1+z}=t_\mathrm{lab}-\frac{R\Tilde{\mu}}{c},
        \label{eq_EATS}
\end{equation}
where $\Tilde{\mu}\equiv \cos\Tilde{\theta}$, $\Tilde{\theta}=\arccos(\hat{r}\cdot\hat{n})$ represents an angle from the line-of-sight (LOS), $z$ denotes redshift of the source, and $t_\mathrm{lab}$ is the lab frame time.
The lab frame time $t_\mathrm{lab}$ can be expressed as an integral \citep{Gill_2018} 
\begin{equation}
    t_\mathrm{lab}=\frac{R_\Delta}{c}\int_0^x \frac{\mathrm{d}x'}{\beta(x')},
        \label{eq_t}
\end{equation}
where $\beta=\sqrt{1-\Gamma^{-2}}$ is dimensionless velocity and $x\equiv R/R_\Delta$ consistent with previous paragraph.

For a structured jet, we divide the jet structure into $N$ segments along the polar angle $\theta$ direction in spherical coordinates, and each segment can be approximate to a uniform ring. 
Similar to the calculation in \citet{Pang_2024}, using the isotropic co-moving spectral luminosity $L_{\nu',\mathrm{iso}}'$, the spectral flux of a uniform ring with an middle half-opening angle $\theta_n$ and an angular width $\Delta\theta$ can be calculated as
\begin{equation}
    \Delta F_{\nu,n}(\theta_n, t_\mathrm{obs}) = \frac{1+z}{4\pi d_L^2} \int_\mathrm{EATS} \Tilde{\delta}_D^3 \frac{L_{\nu',\mathrm{iso}}'(\theta_n, x)}{4\pi}\mathrm{d}\Tilde{\Omega}, 
        \label{eq_delta_flux}
\end{equation}
where $d_L$ represents the luminosity distance, $\Tilde{\delta}_D=[\Gamma(1-\beta\Tilde{\mu})]^{-1}$ stands for the Doppler factor and $\mathrm{d}\Tilde{\Omega}=\mathrm{d}\Tilde{\mu}\mathrm{d}\Tilde{\phi}$ represents the differential solid-angle. 
The integration in the above equation must be carried out over the EATS for a specific $t_\mathrm{obs}$. 
For the observer time $t_\mathrm{obs}$, we have
\begin{equation}
    \mathrm{d}\Tilde{\mu}= \frac{1-\beta\Tilde{\mu}}{\beta x}\mathrm{d}x.
        \label{eq_J}
\end{equation}
Using this expression, the integration in Eq.~\ref{eq_delta_flux} can be performed over $x\in[x_\mathrm{min},x_\mathrm{max}]$ \citep{Gill_2018}.
To find $x_\mathrm{min}$ and $x_\mathrm{max}$, we substitute $\Tilde{\mu}$ with $\Tilde{\mu}_\mathrm{min}$ and $\Tilde{\mu}_\mathrm{max}$ in Eq.~\ref{eq_EATS}, respectively, where $\Tilde{\mu}_\mathrm{min}=\cos(\theta_\mathrm{obs} + \theta_n + \Delta\theta/2)$.
The value of $\mu_\mathrm{max}$ depends on the relative position of LOS and the uniform ring, that is,
\begin{equation}
    \mu_\mathrm{max} = \begin{cases}
        \cos(\theta_\mathrm{obs} - \theta_\mathrm{outer}), &\theta_\mathrm{obs}>\theta_\mathrm{outer}\\
        1, &\theta_\mathrm{outer}\ge\theta_\mathrm{obs}\ge\theta_\mathrm{inner}\\
        \cos(\theta_\mathrm{inner} - \theta_\mathrm{obs}), &\theta_\mathrm{inner}>\theta_\mathrm{obs}
    \end{cases}
\end{equation}
where $\theta_\mathrm{outer}\equiv \theta_n+\Delta\theta/2$ and $\theta_\mathrm{inner}\equiv\theta_n-\Delta\theta/2$.

To integrate over the azimuthal angle $\Tilde{\phi}$, we simplify the calculation by assuming, without loss of generality, that the line of sight (LOS) lies in the $\Tilde{x}O\Tilde{z}$ plane (i.e., $\phi=0$ plane). 
With the coordinate transformation from $(x,y,z)$ to $(\Tilde{x},\Tilde{y},\Tilde{z})$, we have
\begin{equation}
    \cos\Tilde{\phi} = \frac{\Tilde{\mu}\cdot\cos\theta_\mathrm{obs}-\cos\theta}{\sqrt{1-\Tilde{\mu}^2}\cdot\sin\theta_\mathrm{obs}},
    \label{eq_phi}
\end{equation}
where $\theta$ is the polar angle in the $(x,y,z)$ coordinate system. 
As shown in Fig.~\ref{fig1}, when fixing $\Tilde{\theta}$ (which is equivalent to fixing $\Tilde{\mu}$ or $\xi$), the cone with a half-opening angle $\Tilde{\theta}$ around the LOS intersects with the inner and outer edges of uniform ring at $\Tilde{\phi}_\mathrm{inner}$ and $\Tilde{\phi}_\mathrm{outer}$, respectively.
These angles can be easily calculated by substituting $\theta$ for $\theta_n - \Delta\theta/2$ and $\theta_n + \Delta\theta/2$ in Eq.~\ref{eq_phi}.
Using $\Tilde{\phi}_\mathrm{inner}$ and $\Tilde{\phi}_\mathrm{outer}$, we can get the range of $\Tilde{\phi}$
and value of $\int_{\text{Range of $\Tilde{\phi}$} }\mathrm{d}\Tilde{\phi}$ \citep{Granot_2005}.
Finally, the total spectral flux of a structured jet is
\begin{equation}
    F_\nu(t_\mathrm{obs}) = \sum_n^N \Delta F_{\nu, n}(\theta_n , t_\mathrm{obs}).
\end{equation}

The proper motion of the flux centroid, indicative of superluminal motion in certain GRBs \citep{Mooley2018,Mooley2022}, is calculated by considering the projected image of the outflow on the plane $\Tilde{x}O\Tilde{y}$ with coordinates $(\Tilde{x},\Tilde{y})$.
In this plane, the line connecting the LOS to the jet axis aligns with the $\Tilde{x}$-axis (see Fig.~\ref{fig1}).
Due to the system's symmetry,  the flux centroid moves exclusively along the  $\Tilde{x}$-axis, and its position is given by
\begin{equation}
    \Tilde{x}_F(t_\mathrm{obs}) = \frac{\displaystyle\int_0^{\theta_\mathrm{j}}\mathrm{d}\theta~\frac{1+z}{4\pi d_L^2}\int_\mathrm{EATS}L_{\nu'}'R\cos\Tilde{\theta}\cos\Tilde{\phi}~\mathrm{d}\Tilde{\Omega}}{F_\nu(t_\mathrm{obs})}.
    \label{eq_fc}
\end{equation}
The displacement angle $\Tilde{\theta}_F$ the GRB central source's position is determined by $\Tilde{\theta}_F=\Tilde{x}_F/d_A$, where $d_A$ represents the angular diameter distance.
Subsequently, the average apparent velocity of the flux centroid over a time interval $[t_{\mathrm{obs},i}, t_{\mathrm{obs},j}]$ can be computed by
\begin{equation}
    \bar{v}_{\mathrm{app}}=\frac{\Tilde{x}_{c,j}-\Tilde{x}_{c,i}}{t_{\mathrm{obs},j}-t_{\mathrm{obs},i}}.
\end{equation}

In the context of GRBs, the jet structure pertains to characteristics like the opening angle, Lorentz factor, and energy distribution within the relativistic jet.
Typically, GRB jets are simplified to have a homogeneous jet structure (or "top-hat").
In this model, energy per unit solid angle $\varepsilon$ and the initial bulk Lorentz factor $\Gamma_0$ are uniform in jet opening angle, represented by
\begin{equation}
    \varepsilon(\theta) = \frac{E_\mathrm{iso}}{4\pi},\quad \Gamma_0(\theta) = \Gamma_0,\quad 0\le\theta\le\theta_\mathrm{j},
    \label{TH_jet}
\end{equation}
where $E_\mathrm{iso}$ is the isotropic equivalent energy.

Other than the top-hat jet model, the energy distribution $\varepsilon(\theta)\equiv E_\mathrm{iso}/(4\pi)$ and the initial bulk Lorentz factor $\Gamma_0(\theta)$ vary with angle from the jet axis in a structured jet.
Three alternative jet structures are commonly discussed \citep{Lamb_2017}.

(i) For a two-component jet \citep{Zhang_2004,Peng_2005},
\begin{subequations}
    \begin{align}
        \varepsilon(\theta) &= \begin{cases}
        \varepsilon_\mathrm{c} \quad &0\le\theta\le\theta_\mathrm{c},\\
        \varepsilon_\mathrm{w} \quad &\theta_\mathrm{c}<\theta\le\theta_\mathrm{j},
    \end{cases}
    \label{eq_2C_E}
    \\
    \Gamma_0(\theta) &= \begin{cases}
        \Gamma_{0,\mathrm{c}}\quad &0\le\theta\le\theta_\mathrm{c},\\
        \Gamma_{0,\mathrm{w}}\quad &\theta_\mathrm{c}<\theta\le\theta_\mathrm{j},
    \end{cases}
    \label{eq_2C_Lorentz}
\end{align}
\end{subequations}
where $\theta_\mathrm{c}$ is the opening angle of the jet core (or narrow jet), and subscript "c" indicates the jet core parameter, while subscript "w" indicates the parameters in the jet wing  (or a wide jet).

(ii) For a power-law structured jet \citep{Dai_2001,Kumar_2003,Beniamini_2020}, 
\begin{subequations}
    \begin{align}
        \varepsilon(\theta) &= \begin{cases}
        \varepsilon_\mathrm{c} \quad &0\le\theta\le\theta_\mathrm{c},\\
        \varepsilon_\mathrm{c}\Theta^{-a} \quad &\theta_\mathrm{c}<\theta\le\theta_\mathrm{j},
    \end{cases}
    \label{eq_PL_E}
    \\
    \Gamma_0(\theta)-1 &= \begin{cases}
        \Gamma_{0,\mathrm{c}}-1\quad &0\le\theta\le\theta_\mathrm{c},\\
        (\Gamma_{0,\mathrm{c}}-1)\Theta^{-b} \quad &\theta_\mathrm{c}<\theta\le\theta_\mathrm{j},
    \end{cases}
    \label{eq_PL_Lorentz}
\end{align}
\end{subequations}
where $\Theta\equiv\left[1+(\theta/\theta_\mathrm{c})^2\right]^{1/2}$, and $a$,$b$ are the energy angular slope and initial bulk Lorentz factor angular slope of jet, respectively.

And (iii) for a Gaussian jet \citep{Zhang_2002,Kumar_2003},
\begin{equation}
    \varepsilon(\theta) = \varepsilon_\mathrm{c}\exp{\left(-\frac{\theta^2}{2\theta_\mathrm{c}^2}\right)},\quad 
    \Gamma_0(\theta) = \Gamma_{0,\mathrm{c}}\exp{\left(-\frac{\theta^2}{2\theta_\mathrm{c}^2}\right)}.
    \label{eq_G_Jet}
\end{equation}

\begin{figure*}
    \centering
    \includegraphics[width=1.0\textwidth, angle=0]{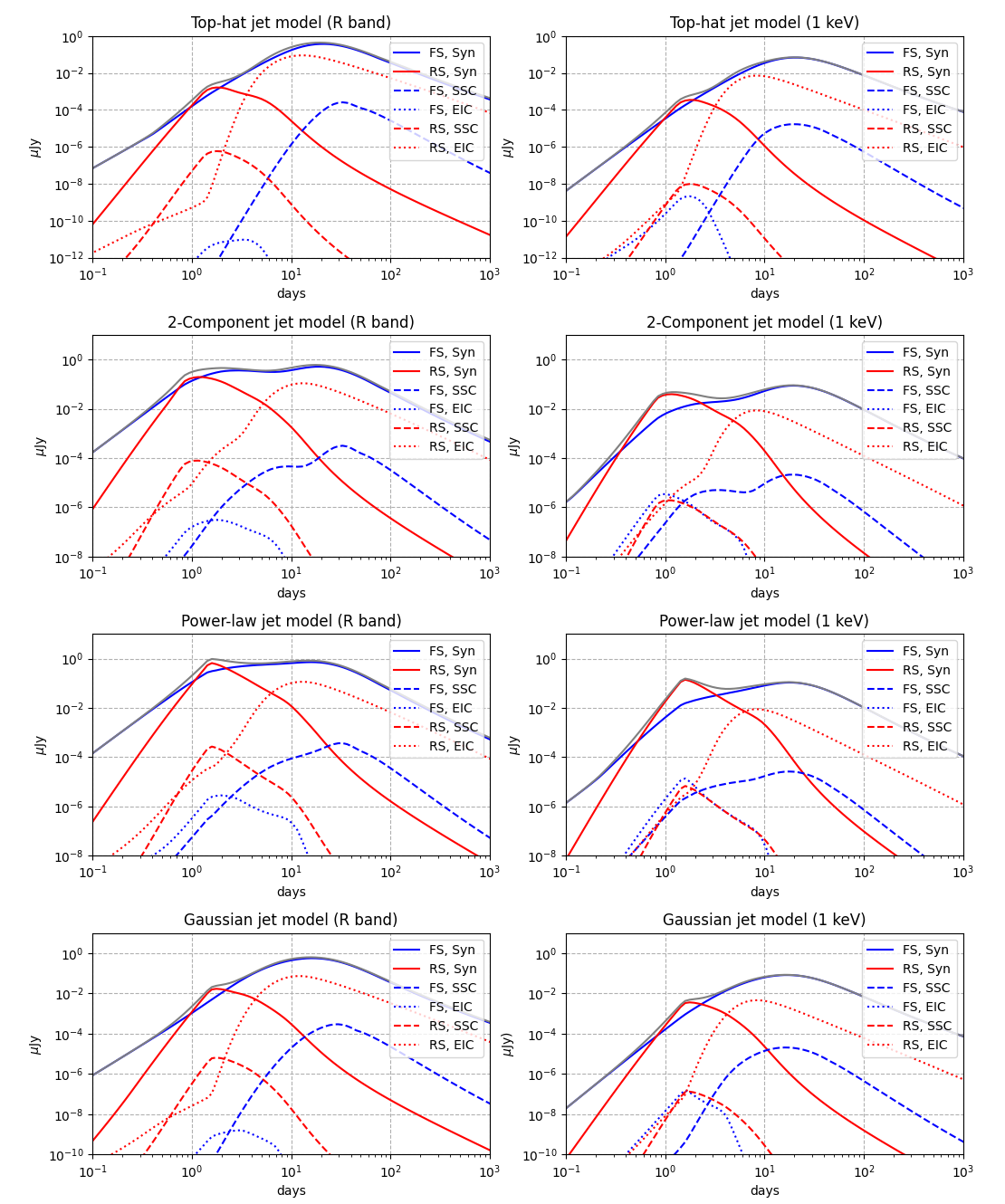}
    \caption{R-band and 1 keV afterglow light curves for different jet models viewed at $20^\circ$. Jets have an isotropic energy $E_{\mathrm{iso,c}}=2\times10^{52}$erg and bulk Lorentz $\Gamma_{0,\mathrm{c}}=300$ in the jet core. The CBM density is $n_1 = 0.1~\mathrm{cm}^{-3}$ ($k=0$). 
    The model values used in each case are: $\theta_\mathrm{c}=\theta_\mathrm{j}=3^\circ$ for a homogeneous (top-hat) jet, $\theta_\mathrm{c}=6^\circ$, $\theta_\mathrm{j}=15^\circ$ for a two-component jet with an isotropic energy $E_{\mathrm{iso,w}}=10^{51}$ erg and bulk Lorentz factor $\Gamma_{0,\mathrm{w}}=10$ in a wide jet, $\theta_\mathrm{c}=3^\circ$, $\theta_\mathrm{j}=10^\circ$ and angular slope $a=b=2.5$ for power-law jet, and (bottom right)$\theta_\mathrm{c}=3^\circ$ for a Gaussian jet with an opening angle $\theta_\mathrm{j}=5.2^\circ$ to keep the relativistic jet. In each subplot, the gray solid lines denotes the total flux, the red and blue solid lines denote the RS and FS synchrotron light curves. The red and blue dashed-lines denote the SSC radiation from RS and FS. The red and blue dotted-lines represent the IC emissions of scatterings of forward shock photons on reverse shocked electrons and reverse shock photons on forward shocked electrons.
    }
    \label{LC1}
\end{figure*}

\begin{figure*}
    \centering
    \includegraphics[width=1.0\textwidth]{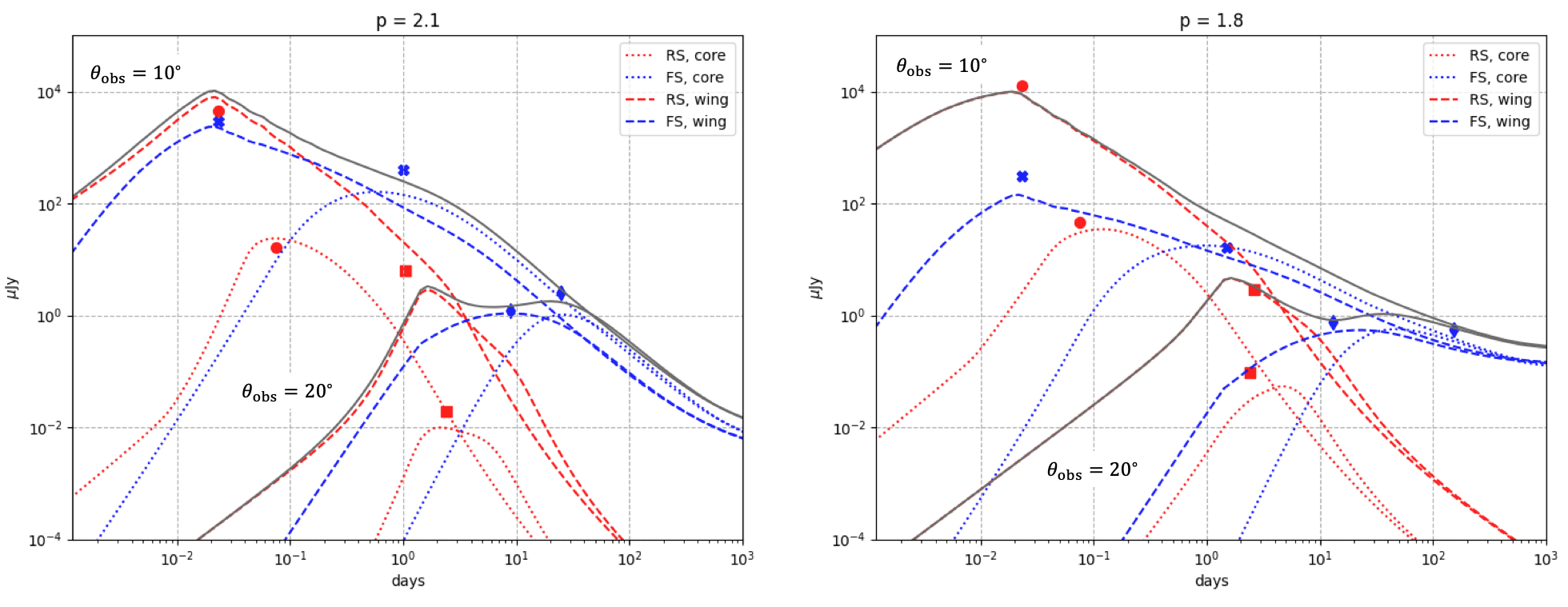}
    \caption{R-band afterglow light curves for power-law jets viewed from $10^\circ$ and $20^\circ$.
    The red lines illustrate the contributions from the RS, while the blue lines represent the FS light curves.
    Dotted lines indicate the jet core components, whereas dashed-lines denote the power-law wing components, with gray solid lines reflecting the total flux. The dots on the plot are the approximated locations of FS and RS peaks from jet core and power-law wings.
    The model parameters employed in these calculations include a half-opening angle $\theta_\mathrm{c} = 3^\circ$ for the core and $\theta_\mathrm{j}=10^\circ$ for the power-law wings.
    The isotropic energy of core is $E_\mathrm{iso,c}=2\times 10^{52}~\mathrm{erg}$, with an initial bulk Lorentz factor is $\Gamma_{0,\mathrm{c}} = 300$.
    The angular slops are set to $a=b=2.5$, and CBM density is $n_1 = 0.1~\mathrm{cm}^{-3}$ ($k=0$).
    }
    \label{LC2}
\end{figure*}

\begin{figure}
    \centering
    \includegraphics[width=1.0\linewidth]{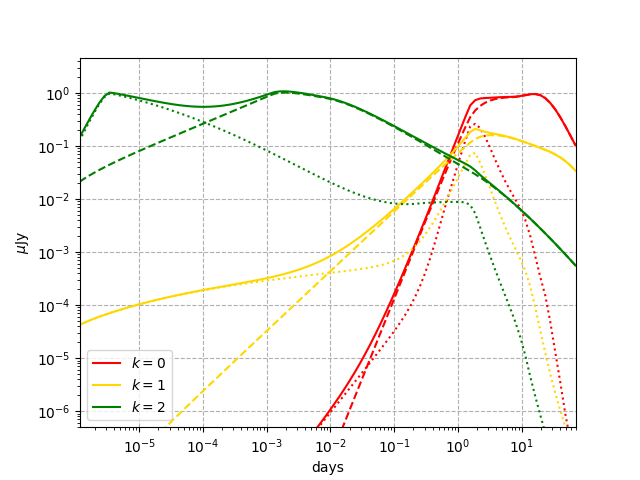}
    \caption{R-band afterglow light curves for a power-law jet in different wind environments. The dash-lines represent the FS emission, while the dotted-lines denote the RS emission.
    The model parameters used are as follows: the half-opening angle of jet core and jet wing are $\theta_\mathrm{c}=3^\circ$ and $\theta_\mathrm{j}=10^\circ$, respectively. 
    The isotropic energy of the jet core is $E_\mathrm{c,iso}=2\times10^{52}\mathrm{erg}$ and initial Lorentz factor is $\Gamma_{0,\mathrm{c}}=300$. 
    The angular slops are set to $a=b=2.5$.}
    \label{fig:wind_type}
\end{figure}

The examples of afterglow light curves for top-hat jet model and three structured jet models mentioned above, from a jet located at 200 Mpc and viewed at an angle of $20^\circ$ are shown in Fig.~\ref{LC1}. 
The light curves in Fig.~\ref{LC2} show the power-law jet viewed at $10^\circ$ and $20^\circ$ for $p=2.1$ and $p=1.8$.
In our examples, we assume that the fraction of internal energy carried by electrons in the shocked regions are $\epsilon_{\mathrm{e},2}=\epsilon_{\mathrm{e},3}=0.1$, while the fraction carried by magnetic fields is $\epsilon_{B,2}=0.01$ for the forward-shocked region and $\epsilon_{B,3}=0.1$ for the reverse-shocked region. This results in a magnetization parameter, as defined by \citet{Harrison_2013}, of $R_B\equiv \epsilon_{B,3}/\epsilon_{B,2}=10\gg1$, which leads to more pronounced RS emission.
There are 4 components in each group of light curves: (i) the FS emission in power-law wings, (ii) the RS emission in power-law wings, (iii) the FS emission in the jet core, and (iv) the RS emission in the jet core. 

The light curves in Fig.~\ref{fig:wind_type} show the emission from power-law jets in stratified environments, observed from an angle of $\theta_\mathrm{obs}=20^\circ$. For a comparison, we set $n_1(R=10^{17}\mathrm{cm})\equiv 0.1 \mathrm{cm}^{-3}$ for different $k$-values.
In stratified environments, the CBM can be very dense at early times, when the shocks are still at small radii, causing the self-absorption frequency $\nu_\mathrm{a}'$ to exceed both $\nu_\mathrm{m}'$ and $\nu_\mathrm{c}'$. During this period, the spectrum can be described as \citep{Wu_2003, Li_&_Gao_&_Ai_2023}
\begin{equation}
    \frac{P_{\nu'}'}{P_{\nu',\mathrm{max}}'} = \begin{cases}
        \left(\frac{\nu'}{\nu_\mathrm{m}'}\right)^2\left(\frac{\nu_\mathrm{m}'}{\nu_\mathrm{a}'}\right)^{\frac{p+4}{2}}\left(\frac{\nu_\mathrm{a}'}{\nu_\mathrm{c}'}\right)^{-\frac{1}{2}}, &\nu'<\nu_\mathrm{m}'\\
        \left(\frac{\nu'}{\nu_\mathrm{a}'}\right)^{5/2}\left(\frac{\nu_\mathrm{a}'}{\nu_\mathrm{c}'}\right)^{-\frac{p}{2}}
        \left(\frac{\nu_\mathrm{c}'}{\nu_\mathrm{m}'}\right)^{\frac{1-p}{2}}, &\nu_\mathrm{m}'<\nu'<\nu_\mathrm{a}'\\
        \left(\frac{\nu'}{\nu_\mathrm{c}'}\right)^{-p/2} \left(\frac{\nu_\mathrm{c}'}{\nu_\mathrm{c}'}\right)^{\frac{1-p}{2}}, &\nu_\mathrm{a}'<\nu'
    \end{cases}
    \label{eq_wind_type_syn1}
\end{equation}
for slow cooling, and
\begin{equation}
    \frac{P_{\nu'}'}{P_{\nu',\mathrm{max}}'}=\begin{cases}
        \left(\frac{\nu'}{\nu_\mathrm{c}'}\right)^2\left(\frac{\nu_\mathrm{c}'}{\nu_\mathrm{a}'}\right)^3\left(\frac{\nu_\mathrm{a}'}{\nu_\mathrm{m}'}\right)^{\frac{1-p}{2}}, &\nu'<\nu_\mathrm{c}'\\
        \left(\frac{\nu'}{\nu_\mathrm{a}'}\right)^{5/2}\left(\frac{\nu_\mathrm{a}'}{\nu_\mathrm{m}'}\right)^{-\frac{p}{2}}\left(\frac{\nu_\mathrm{m}'}{\nu_\mathrm{c}'}\right)^{-\frac{1}{2}}, &\nu_\mathrm{c}'<\nu'<\nu_\mathrm{a}'\\
        \left(\frac{\nu'}{\nu_\mathrm{m}'}\right)^{-p/2}\left(\frac{\nu_\mathrm{m}'}{\nu_\mathrm{c}'}\right)^{-\frac{1}{2}}, &\nu_\mathrm{a}'<\nu'
    \end{cases}
    \label{eq_wind_type_syn2}
\end{equation}
for fast cooling.
In Fig.~\ref{fig:wind_type}, the RS emission is stronger in stratified environments at early times due to the higher CBM density at small radii, which leads to a stronger RS in the jet. A larger external density power-law index $k$ leads to a shallower rise toward the peak of the light curve, as the mass swept by the shell increases more slowly ($\propto R^{3-k}$). In the wind environment ($k=2$), the RS emission peaks at early times due to the self-absorption frequency $\nu_\mathrm{a}$ decreasing and passing through the observation frequency. The break in RS emission at $\sim1~\mathrm{day}$ is caused by the RS crossing the ejecta shell.

If an observation angle $\theta_\mathrm{obs}$ satisfies $\theta_\mathrm{c}<\theta_\mathrm{obs}<\theta_\mathrm{j}$, the emission of jet wings can be approximated as on-axis emission from the forward-reverse shocks with a low bulk Lorentz factor ($\Gamma_0\sim 10^1$). The peak time ($k=0$) can be approximated as \citep{Zhang_Ze_lin_2022}
\begin{equation}
    t_\Delta \approx 0.668 ~\Gamma_{0,1}^{-8/3}E_{\mathrm{iso},51}^{1/3}n_{0,0}^{-1/3}~\text{days},
    \label{cross_time}
\end{equation}
where $\Gamma_0$ is the Lorentz factor of the jet wing at $\theta=\theta_\mathrm{obs}<\theta_\mathrm{j}$, $E_\mathrm{iso}\equiv4\pi\varepsilon(\theta)$ is the isotropic energy of the jet wing at $\theta=\theta_\mathrm{obs}$
and $n_0$ is the number density of ISM.
The convention $Q=10^n Q_n$ is adopted for cgs unites in this Section.

At the crossing time $t_\Delta$, the typical frequencies and peak flux density of forward and reverse shock synchrotron emission are \citep{Kobayashi_2000,Dai_&_Cheng_2001,Shao_2005, Yi_2013}
\begin{align}
    \begin{split}
        \nu_{\mathrm{m},f}(t_\Delta) &\approx 9.84 \times 10^{11} \left(\frac{p-2}{p-1}\right)^2 \\
        &\quad\times\epsilon_{\mathrm{e},f,-1}^2\epsilon_{B,f,-2}^{1/2} \Gamma_{0,1}^4 n_{0,-1}^{1/2}~\text{Hz},
    \end{split}\\
    \begin{split}
        \nu_{\mathrm{m},r}(t_\Delta) & \approx 1.04\times 10^{9} \left(\frac{p-2}{p-1}\right)^2 \\
        &\quad \times \epsilon_{\mathrm{e},r,-1}^2\epsilon_{B,r,-2}^{1/2}\Gamma_{0,1}^2 n_{0,-1}^{1/2}~\text{Hz},
    \end{split}
\end{align}
for $p>2$, and
\begin{align}
    \begin{split}
        \nu_{\mathrm{m},f}(t_\Delta) &\approx 4.47\times 10^{6} \\
        &\quad \times \epsilon_{\mathrm{e},f,-1}^{\frac{2}{p-1}}\epsilon_{B,f,-2}^{\frac{1}{2(p-1)}} \Gamma_{0,1}^{\frac{p+2}{p+1}} n_{0,-1}^{\frac{1}{2(p-1)}}~\text{Hz},
    \end{split}\\
    \begin{split}
        \nu_{\mathrm{m},r}(t_\Delta) &\approx 9.51\times 10^{7}\\
        &\quad \times \epsilon_{\mathrm{e},r,-1}^{\frac{2}{p-1}}\epsilon_{B,r,-2}^{\frac{1}{2(p-1)}}\Gamma_{0,1}^{\frac{p}{p-1}}n_{0,-1}^{\frac{1}{2(p-1)}}~\text{Hz}
    \end{split}
\end{align}
for $p=1.8$. 
The observed peak flux is given by
\begin{align}
    \begin{split}
        F_{\nu,\mathrm{m},f}(t_\Delta)&\approx 24.7~\mu\text{Jy} \\
        &\quad\times d_{L,28}^{-2}\epsilon_{B,f,-2}^{1/2}E_{\mathrm{iso},51}n_{0,-1}^{1/2},
    \end{split}\\
    \begin{split}
        F_{\nu,\mathrm{m},r}(t_\Delta)&\approx 75.8~\mu\text{Jy}\\
        &\quad\times d_{L,28}^{-2} \epsilon_{B,r,-2}^{1/2}\Gamma_{0,1} E_{\mathrm{iso},51} n_{0,-1}^{1/2},
    \end{split}
\end{align}
where the subscript $f$ and $r$ represent forward and reverse shocks, respectively.

If a top-hat jet (or jet core) with a half-opening angle $\theta_\mathrm{c}$ can be approximated as a point source (which means $\theta_\mathrm{obs}\gg\theta_\mathrm{c}$), the jet can be described using a single Doppler factor. 
The afterglow light curve can then be calculated by
\begin{equation}
    F_\nu(\theta_\mathrm{obs},t) = a^3 F_{\nu/a}(\theta_\mathrm{obs}=0,at) ,
\end{equation}
where $a$ is the ratio of the Doppler factor for an off-axis observer to that of an on-axis observer. 
When the RS crosses the shell, the RS emission reaches its peak, and the peak time of RS can be approximated by \citep{Pang_2024}
\begin{equation}
    t_{\mathrm{p},r}= \left[ 1 + \Gamma_0^2(\theta_\mathrm{obs}-\theta_\mathrm{c})^2 \right]l/(\Gamma_0^{8/3}c)
    \label{eq_tpr}
\end{equation}\\
where $l=[ 3E_\mathrm{iso}/(4\pi n_0m_\mathrm{p}c^2) ]^{1/3}$ denotes Sedov length.
The FS emission reaches its peak at $t_{\mathrm{p},f}$, and it determined by
\begin{equation}
    \theta_\mathrm{obs} - \theta_\mathrm{c} = \frac{1}{\Gamma(\xi=\xi_f)}, 
\end{equation}
and substituting $\xi_f$ into the EATS (Eq.~\ref{eq_EATS}) yields the FS peak time $t_{\mathrm{p},f}$. 
For $p>2$, the peak flux of FS can be approximated as \citep{Nakar_2002,Lamb_2017}
\begin{equation}
    \begin{split}
        F_{\nu,\mathrm{p},f}&\approx 155 \frac{g_1(p)}{g_1(2.2)} d_{L,28}^{-2} \epsilon_{\mathrm{e},f,-1}^{p-1}\epsilon_{B,f,-2}^{(p+1)/4} n_{0,0}^{(p+1)/4}\\
        &\quad\times \frac{E_{\mathrm{iso}}}{10^{50.7}\text{erg}}\left(\frac{\nu_{\mathrm{obs}}}{10^{14.7}\text{Hz}}\right)^{(1-p)/2}\theta_{\mathrm{obs},-1}^{-2p}\theta_\mathrm{c}^2 \quad\mu\text{Jy},
    \end{split}
    \label{eq_Fpf1}
\end{equation}
where we used the approximation $(1-\cos\theta_\mathrm{c})\approx \theta_\mathrm{c}^2/2$ in the above equation, and $g_1(p)=10^{-0.31p}(p-0.04)[(p-2)/(p-1)]^{p-1}$.
For $1<p<2$, the peak flux of FS can be derived similarly, as done by \citet{Nakar_2002}, and is given by
\begin{equation}
    F_{\nu<\nu_\mathrm{c},p,f}=\left(\frac{\nu_{\mathrm{obs}}}{\nu_{\mathrm{m},f}(t_{\mathrm{p},f})}\right)^{(1-p)/2}F_{\nu,\mathrm{m},f}(R),
    \label{eq_Fpf2}
\end{equation}
where the radius of shock is approximately $R\approx E_{\mathrm{iso}}/(4\pi\Gamma^2 n_1m_\mathrm{p}c^2)$.
We can use Eq.~\ref{eq_tpr} - \ref{eq_Fpf2} to estimate the emission from an off-axis jet core, and the approximate value of the jet core peaks is shown in Fig.~\ref{LC2}.

\subsection{Mixed jet model} \label{subsec:Mix_jet}
To explain the broad-band observations of GRB 221009A, \citet{Zhang_2024} provided a physical picture involving two jet components: a Poynting-flux-dominated narrow jet surrounded by a matter-dominated structured jet wing (referred to as Mixed jet model).
An ultra-relativistic jet launched from the GRB central engine may be Poynting-flux-dominated, characterized by a high magnetization parameter $\sigma$, defined as the Poynting-to-kinetic flux ratio \citep{Zhang_2005}:
\begin{equation}
    \sigma\equiv \frac{B_4'^2}{4\pi n_4' m_\mathrm{p}c^2},
    \label{eq_sigma}
\end{equation}
where $B_4'$ and $n_4'$ are the magnetic field strength and number density of the unshocked jet in the rest frames of fluids.
As the jet propagates through and breaks out of the progenitor vestige (the stellar envelope of the collapsar or ejecta cloud of the compact binary merger), the material of the jet and cocoon is rearranged into a inhomogeneous shell, potentially forming a mixed jet \citep{Salafia2022}.

Assuming the mixed jet consists of a homogeneous jet core surrounded by a matter-dominated power-law wing, the energy per unit solid angle $\varepsilon(\theta)$ and the angle-dependent initial Lorentz factor are defined as follows:
\begin{subequations}
    \begin{align}
        \varepsilon(\theta) &= \begin{cases}
        \varepsilon_\mathrm{c}, & 0\le\theta\le\theta_\mathrm{c},\\
        \varepsilon_\mathrm{c}\Theta^{-a}, & \theta_\mathrm{c}<\theta\le\theta_\mathrm{j},
        \end{cases}
        \label{eq_Mix_jet_E}
    \end{align}
    \begin{align}
        \Gamma_0(\theta)-1=\begin{cases}
        \Gamma_{0,\mathrm{c}}-1, &0\le\theta\le\theta_\mathrm{c},\\
        (\Gamma_{0,\mathrm{w}}-1)\Theta^{-b}, &\theta_\mathrm{c}<\theta\le\theta_\mathrm{j},
    \end{cases}
    \label{eq_Mix_jet_Gamma}
    \end{align}
\end{subequations}
where $\Theta$, $a$ and $b$ are defined as in the power-law structured jet, and $\Gamma_{0,\mathrm{c}}>\Gamma_{0,\mathrm{w}}$ reflects the impulsive acceleration of the high-$\sigma$ jet core \citep{Granot_2011}.

\begin{figure}
    \centering
    \includegraphics[width=1.0\linewidth]{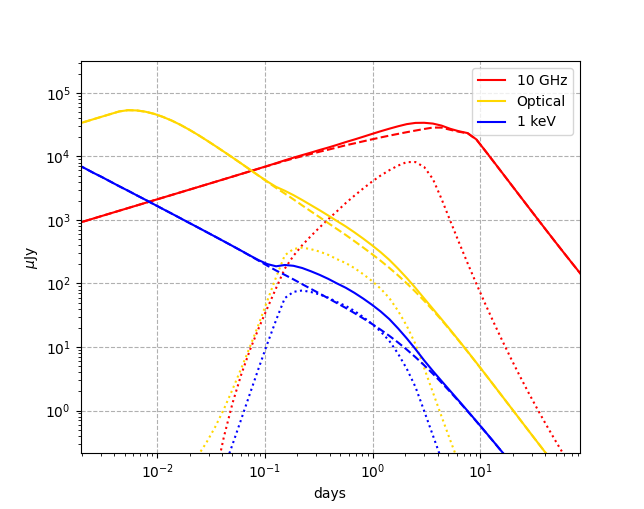}
    \caption{Afterglow light curves for a mixed jet viewed from slightly off-axis.
    The dashed lines denote the FS emission, the dotted lines represent RS emission, and solid lines is the total flux.
    The values of parameters are as follows: $\theta_\mathrm{obs}=2^\circ$, $\theta_\mathrm{c}=5^\circ$, $\theta_{\mathrm{j}} = 10^\circ$, $\Gamma_{0,\mathrm{c}}=300$, $\Gamma_{0,\mathrm{w}}=100$, $E_\mathrm{c,iso}=2\times10^{52}$, $a=b=2.5$, and the magnetization parameter is set to $\sigma=10$.
    }
    \label{fig:Mixed_jet}
\end{figure}

In a Poynting-flux-dominated jet core with high-$\sigma$, a RS forms when the total pressure in the forward-shocked region exceeds the magnetic pressure in the unshocked jet during the shell's coasting phase ($R<R_\mathrm{dec}\sim (1+\sigma)^{-1/3}l/\Gamma_0^{2/3}$). This leads to a constraint on $\sigma$ \citep{Zhang_2005, Ma_2022}
\begin{equation}
    \sigma < 100\left(\frac{\Gamma_0}{300}\right)^4\left(\frac{T}{10~\mathrm{s}}\right)^{3/2}\left(\frac{E_\mathrm{iso}}{10^{52}~\mathrm{ergs}}\right)^{-1/2},
    \label{ineq_sigma}
\end{equation}
where $T$ is the duration of GRB central engine activity.
When the mixed jet is mildly inclined ($\theta_\mathrm{obs}\gtrsim \theta_\mathrm{j}$) with high-$\sigma$ in the core, the absence of RS emission dose not significantly affect the light curve, as the emission from forward and reverse shocks in the jet wing dominants early afterglows. 
However, when the jet is slightly off-axis ($\theta_\mathrm{obs}\lesssim\theta_\mathrm{c}$), it exhibits more distinctive features, as shown in Fig.~\ref{fig:Mixed_jet}.
The early optical and X-ray afterglow are dominated by the FS emission from the jet core, and the bright RS emission contributes prominently to the light curves during their decay.
Additionally, the radio emission from FS in the jet core peaks later, while the RS emission may produce a distinct feature before the FS peak time.

\section{Fit to GRB 170817A/GW170817} \label{sec:results}
GRB 170817A/GW170817 stands out as the first binary neutron star merger detected in gravitational waves, and notably, it was also accompanied by electromagnetic radiation. 
The comprehensive study by \citet{Makhathini_2021} involved collecting and reprocessing radio, optical, and X-ray data spanning from 0.5 days to 1231 days post-merger.
Subsequent reports and analyses of observations in the X-ray band at very late times ($>10^3$ days) were provided by \citet{Troja_2021} and \citet{OConnor_2022}. 
Additionally, \citet{Mooley2018,Mooley2022} demonstrated the superluminal motion of the flux centroid of this burst.
Various hypotheses have been proposed to explain the afterglow of GRB 170817A.
These include scenarios such as a structured jet viewed off-axis \citep{Gill_2018, Lamb_2019_170817}, energy injection into a top-hat jet \citep{Li_2018}, refreshed shocks in a top-hat jet viewed off-axis \citep{Lamb2020}, a relativistic electron-positron wind off-axis \citep{Li_2021}, and RS emission in an off-axis top-hat jet \citep{Pang_2024}.

\begin{deluxetable}{lc}\label{tab:parameters}
\tablecaption{Model parameters for different jet models}
\tablehead{
\colhead{Jet model} & \colhead{Parameters}
}
\startdata
2C & $p$, $n_1$, $\theta_{\mathrm{obs}}$, $\theta_\mathrm{j}$, $\theta_\mathrm{c}$, $\Gamma_{0,\mathrm{c}}$, $E_{\mathrm{iso,c}}$,\\
{   }&$\Gamma_{0,\mathrm{w}}$, $E_{\mathrm{iso,w}}$, $\epsilon_\mathrm{e}$, $\epsilon_{B,2}$, $\epsilon_{B,3}$\\
PL & $p$, $n_1$, $\theta_{\mathrm{obs}}$, $\theta_\mathrm{j}$, $\theta_\mathrm{c}$, $\Gamma_{0,\mathrm{c}}$, $E_{\mathrm{iso,c}}$, $a$, $\epsilon_\mathrm{e}$, $\epsilon_{B,2}$, $\epsilon_{B,3}$\\
G & $p$, $n_1$, $\theta_{\mathrm{obs}}$, $\theta_\mathrm{j}$, $\theta_\mathrm{c}$, $\Gamma_{0,\mathrm{c}}$, $E_{\mathrm{iso,c}}$, $\epsilon_\mathrm{e}$, $\epsilon_{B,2}$, $\epsilon_{B,3}$\\
MJ & $p$, $n_1$, $\theta_\mathrm{obs}$, $\theta_\mathrm{j}$, $\theta_\mathrm{c}$, $\Gamma_{0,\mathrm{c}}$, $E_\mathrm{iso,c}$,\\
{   } &$\Gamma_{0,\mathrm{w}}$, $a$, $\epsilon_e$, $\epsilon_{B,2}$, $\epsilon_{B,3}$, $\sigma$ \\
\enddata
\tablecomments{For simplicity, we have assumed $a=b$ in Eq.~\ref{eq_PL_E}, \ref{eq_PL_Lorentz}, \ref{eq_Mix_jet_E} and \ref{eq_Mix_jet_Gamma} for power-law (PL) jet model and mixed jet (MJ) model.}
\end{deluxetable}

We develop our model in Section \ref{sec:model}. We here use Markov Chain Monte Carlo (MCMC) method to fit the afterglow data of GRB 170817A, which encompasses multi-wavelength afterglow observations and VLBI proper motion data.
Typically, short GRBs like GRB 170817A are associated with the merger of two compact stars, and the CBM is often presumed to be a constant density medium. 
In our fitting procedure, we adopt $k=0$ to accommodate this assumption.
The duration of prompt emission is about 2 seconds, we set the initial width of the shell to $6\times 10^{10}~\mathrm{cm}$.
In line with prior investigations \citep{ZKP03,Harrison_2013,Gao_2015}, we maintain the assumption that the electron equipartition parameter $\epsilon_e$ and the electron power-law index $p$ remain consistent across both the forward- and reverse-shocked regions.  
Nevertheless, considering that the outflow may feature an initial magnetic field originating from the central engine, we introduce the notion that two shock regions possess different magnetic equipartition parameters $\epsilon_B$.

Different jet models require different sets of parameters.
The required parameters for four structured jet models (i.e., two-component (2C) model, power-law (PL) model, Gaussian (G) model, and mixed jet (MJ) model), are shown in Table~\ref{tab:parameters}.
In this paper, we apply uniform priors for the following parameters: $p$, $\log n_1$, $\theta_\mathrm{obs}$, $\theta_\mathrm{j}$, $\theta_\mathrm{c}$, $\log \Gamma_{0,\mathrm{c}}$, $\log E_\mathrm{iso,c}$, $\log \Gamma_{0,\mathrm{w}}$, $\log E_\mathrm{iso,w}$, $a$, $\log \epsilon_e$, $\log\epsilon_{B,2}$, $\log \epsilon_{B,3}$ and $\log \sigma$.
The specific ranges for these priors are shown in Table \ref{tab:prior}.
For GRB 170817A, the binary inclination angle $\theta_{JN}={151_{-11}^{+15}}^\circ$ was determined by \citet{Abbott_2019} through joint gravitational-wave (GW) and electromagnetic (EM) observations. 
Therefore, we confine the prior range for $\theta_\mathrm{obs}$ to $0.23\le \theta_\mathrm{obs} \le 0.7$.

\begin{figure*}[h]
    \centering
    \includegraphics[width=1.0\textwidth, angle=0]{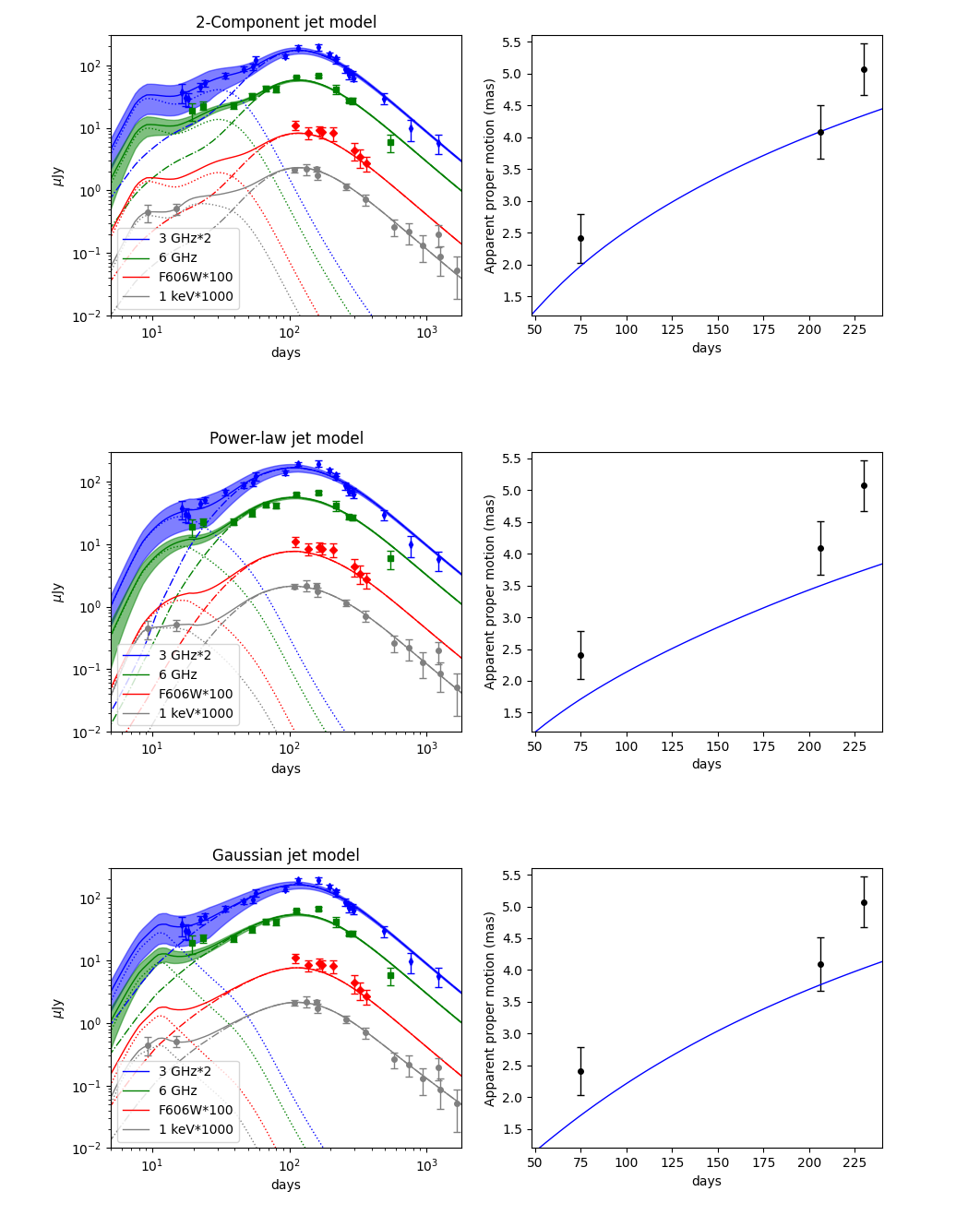}
    \caption{The left column is multi-wavelength afterglow light curves of GRB 170817A with the best-fit curves in our afterglow model. Different colors are used to represent different bands. 
    The RS light curves are depicted as dotted lines, while the FS light curves are represented by dash-dotted lines. 
    The shaded region represent the variability from scintillation.
    The right column is proper motion of GRB 170817A relative to 8 d Hubble Space Telescope (HST) measurement. The blue line represents the result obtained from our model, based on the best-fitting parameters. The proper motion measured between 8 d - 75 d, 8 d - 206 d, and 8 d - 230 d is $2.41\pm0.38$ mas, $4.09\pm0.42$ mas, and $5.07\pm0.40$ mas, respectively \citep{Mooley2022}.
    }
    \label{Fitting}
\end{figure*}
\begin{figure*}
    \centering
    \includegraphics[width=1.0\textwidth, angle=0]{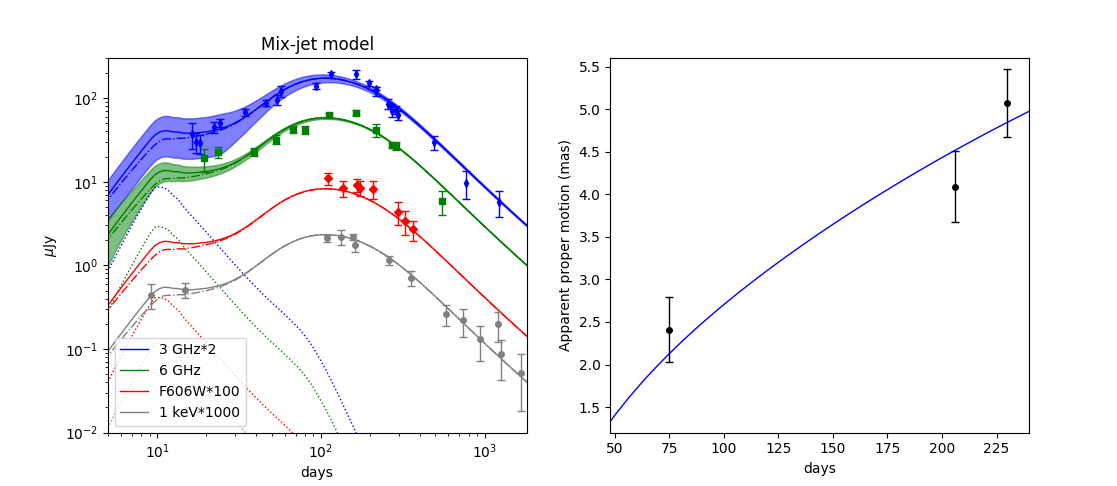}
    \caption{Same as Fig.~\ref{Fitting}, but for the mixed jet model.}
    \label{MJ_fitting}
\end{figure*}

\begin{splitdeluxetable*}{lcccccccBccccccc}\label{tab:prior}
\tablecaption{Priors for model parameters: 1 break}
\tablehead{
\colhead{Jet model} &\colhead{Model parameters}\\
\cline{2-15}
{   } &\colhead{$p$} &\colhead{$\log [n_1(\mathrm{cm}^{-3})]$} &\colhead{$\theta_\mathrm{obs}$(rad)} &\colhead{$\theta_\mathrm{j}$(rad)} &\colhead{$\theta_\mathrm{c}$(rad)} &\colhead{$\log \Gamma_{0,\mathrm{c}}$} &\colhead{$\log [E_\mathrm{iso,c}(\mathrm{erg})]$} &\colhead{$\log \Gamma_{0,\mathrm{w}}$} &\colhead{$\log [E_\mathrm{iso,w}(\mathrm{erg})]$} &\colhead{$a$} &\colhead{$\log \epsilon_e$} &\colhead{$\log\epsilon_{B,2}$} &\colhead{$\log \epsilon_{B,3}$} &\colhead{$\log \sigma$}
}
\startdata
2C &$(2,3)$ &$(-5,1)$ &$(0.23,0.7)$ &$(0,0.7)$ &$(0,0.7)$ &$(1,3)$ &$(49,54)$ &$(0.7,3)$ &$(48,53)$ &-- &$(-7,0.5)$ &$(-7,0.5)$ &$(-7,0.5)$ &-- \\ 
PL &$(2,3)$ &$(-5,1)$ &$(0.23,0.7)$ &$(0,0.7)$ &$(0,0.7)$ &$(1,3)$ &$(49,54)$ &-- &-- &$(0.1,10)$ &$(-7,0.5)$ &$(-7,0.5)$ &$(-7,0.5)$ &--\\
G  &$(2,3)$ &$(-5,1)$ &$(0.23,0.7)$ &$(0,0.7)$ &$(0,0.7)$ &$(1,3)$ &$(49,54)$ &-- &-- &-- &$(-7,0.5)$ &$(-7,0.5)$ &$(-7,0.5)$ &-- \\
MJ &$(2,3)$ &$(-5,1)$ &$(0.23,0.7)$ &$(0,0.7)$ &$(0,0.7)$ &$(1,3)$ &$(49,54)$ &$(0.7,3)$ &-- &$(0.1,10)$ &$(-7,0.5)$ &$(-7,0.5)$ &$(-7,0.5)$ &$(0,3)$\\
\enddata
\end{splitdeluxetable*}

\begin{splitdeluxetable*}{lcccccccBccccccc}\label{tab:fitting_results}
\tablecaption{Best fitting results for model parameters: 1 break}
\tablehead{
\colhead{Jet model} &\colhead{Model parameters}\\
\cline{2-15}
{ } &\colhead{$p$} &\colhead{$\log [n_1(\mathrm{cm}^{-3})]$} &\colhead{$\theta_\mathrm{obs}$(rad)} &\colhead{$\theta_\mathrm{j}$(rad)} &\colhead{$\theta_\mathrm{c}$(rad)} &\colhead{$\log \Gamma_{0,\mathrm{c}}$} &\colhead{$\log [E_\mathrm{iso,c}(\mathrm{erg})]$} &\colhead{$\log \Gamma_{0,\mathrm{w}}$} &\colhead{$\log [E_\mathrm{iso,w}(\mathrm{erg})]$} &\colhead{$a$} &\colhead{$\log \epsilon_e$} &\colhead{$\log\epsilon_{B,2}$} &\colhead{$\log \epsilon_{B,3}$} &\colhead{$\log \sigma$}
}
\startdata
2C &$2.160_{-0.005}^{+0.005}$ &$-2.816_{-0.411}^{+0.321}$ &$0.251_{-0.008}^{+0.017}$ &$0.079_{-0.015}^{+0.012}$ &$0.038_{-0.009}^{+0.008}$ &$2.436_{-0.470}^{+0.325}$ &$53.579_{-0.385}^{+0.284}$ &$1.845_{-0.368}^{+0.345}$ &$52.196_{-0.493}^{+0.354}$ &-- &$-1.135_{-0.527}^{+0.388}$ &$-4.787_{-0.969}^{+1.101}$ &$-2.897_{-1.092}^{+0.883}$ &-- \\
PL &$2.160_{-0.006}^{+0.006}$ &$-3.190_{-0.489}^{+0.358}$ &$0.255_{-0.006}^{+0.005}$ &$0.101_{-0.001}^{+0.002}$ &$0.091_{-0.010}^{+0.004}$ &$2.591_{-0.068}^{+0.049}$ &$53.467_{-0.474}^{+0.352}$ &-- &-- &$4.278_{-1.140}^{+1.413}$ &$-1.308_{-0.533}^{+0.383}$ &$-5.066_{-1.071}^{+1.020}$ &$-2.839_{-0.913}^{+0.797}$ &--\\
G  &$2.159_{-0.005}^{+0.005}$ &$-2.949_{-0.595}^{+0.399}$ &$0.275_{-0.014}^{+0.016}$ &$0.103_{-0.002}^{+0.005}$ &$0.036_{-0.001}^{+0.002}$ &$2.955_{-0.052}^{+0.034}$ &$53.511_{-0.594}^{+0.359}$ &-- &-- &-- &$-1.367_{-0.704}^{+0.534}$ &$-4.123_{-1.261}^{+1.090}$ &$-2.910_{-1.221}^{+1.004}$ &--\\
MJ &$2.158_{-0.005}^{+0.005}$ &$-2.699_{-0.288}^{+0.167}$ &$0.266_{-0.004}^{+0.004}$ &$0.134_{-0.023}^{+0.021}$ &$0.084_{-0.006}^{+0.006}$ &$2.452_{-0.054}^{+0.037}$ &$53.797_{-0.289}^{+0.150}$ &$2.420_{-0.064}^{+0.045}$ &-- &$4.453_{-0.606}^{+0.399}$
&$-2.006_{-0.778}^{+0.798}$ &$-4.988_{-1.204}^{+1.125}$ &$-4.489_{-1.375}^{+1.401}$ &$0.074_{-0.054}^{+0.113}$\\
\enddata
\end{splitdeluxetable*}

In the left column of Fig.~\ref{Fitting} and Fig.~\ref{MJ_fitting}, we present the multi-wavelength observations of GRB 170817A afterglows, along with the fitting results obtained using the FS-RS model for three classical structured jet models and the mixed jet model.
Our afterglow fittings reveal that, in the two-component jet model, the gradual rise can be attributed to the difference in RS peak times between wide and narrow jet. 
In the power-law and Gaussian jet model, this rise is explained by the difference in RS and FS peak times.
In the mixed jet model, however, the RS emission is much weaker than the FS emission, contributing insignificantly to the light curves. This is due to the low magnetization parameter $R_B \equiv \epsilon_{B,3}/\epsilon_{B,2}\approx 3$ in our fitting result. The gradual rise in this case results from the FS emission originating from the jet wing.
Specifically, the RS emission produces a peak at $\sim 10$ days, while the peak time of the FS emission is roughly at $100$ days.
The X-ray excess at late times, approximately 1200 days post burst, is possibly attributed to the emission originating from the central engine \citep{Troja_2021,Hajela_2022}.
In the right column of Fig.~\ref{Fitting} and Fig.~\ref{MJ_fitting}, we present displacement data of the flux centroid and the fitting results based on our model with the fitting parameters.
The alignment between our model and the observed data indicates that our model can effectively reproduce the superluminal motion of the flux centroid. 
Fig.~\ref{corner2C}-\ref{cornerMJ} displays the corner plots showing the results of our MCMC parameter estimation.
The best-fit values of four structured jet models are presented in Table~\ref{tab:fitting_results}.
According to our fitting results, GRB 170817A locates in a low density CBM with number density $n_1\sim 10^{-3}~\mathrm{cm}^{-3}$, which is comparable with \citet{Lamb2020} and \citet{Li_2021}. 
Notably, all four structured jet models yield similar results for the viewing angle and jet half-opening angle. 
The derived viewing angle $\theta_\mathrm{obs}\approx 17^\circ$ is close to an upper limit on binary inclination angle by combining gravitational-wave and electromagnetic constraints \citep{Abbott_2019}. 
The best-fitting parameters for structured jet models suggest a jet half-opening angle $\theta_\mathrm{j}$  around $5^\circ - 7^\circ$, aligning well with the typical values and comparable with \citet{Li_2021}. 
In our fitting results, the half-opening angle of the jet core $\theta_\mathrm{c}\lesssim 5^\circ$, which is consistent with \citet{Mooley2018,Mooley2022}. Comparing to our previous work \citep{Pang_2024}, we have a better fit for the VLBI proper motion of GRB 170817A, a larger viewing angle and jet half-opening angle in fitting results due to the structured jet.
In the mixed jet model, the jet core is moderately magnetized, with $\sigma\approx 1.19$, exceeding the critical value required for the RS that can be formed for a given energy and bulk Lorentz factor.

Scintillation at 3~GHz and 6~GHz is shown as shaded regions in Fig.~\ref{Fitting} and Fig.~\ref{MJ_fitting} representing the maximum and minimum variability.
Here, we only consider strong scintillation when the observation frequencies are lower than transition frequency $\nu_0$.
The modulation index $m$ (the ratio of the standard deviation and mean value of the flux density) can be expressed in terms of the angular size of the first Fresnel zone $\theta_{\mathrm{F}0}$, $\nu_0$ and angular size of the source, as derived by \citet{Granot_van_der_Horst_2014}.
Following \citet{Walker1998, Walker2001}, \citet{Granot_van_der_Horst_2014} and \citet{Lamb_2019_170817}, we use $\theta_{\mathrm{F}0}=6.23\times 10^4~\mathrm{SM}^{0.6}\nu_0^{-2.2}~\mu\mathrm{as}$, where $\mathrm{SM}$ is the scattering measure, typically taken to be $\mathrm{SM}\sim 10^{-3.5}~\mathrm{kpc}/\mathrm{m}^{20/3}$, and $\nu_0\sim 10~\mathrm{GHz}$ \citep{Cordes2002}.
Our fitting results and calculations indicate that the scintillation significantly affect the 3~GHz light curves at early times when the source is compact. However, in 6~GHz band, the reverse shock and structure of the jet are still required to explain the gradual rise at early times.

\section{Discussion and conclusions} \label{sec:4}

\begin{figure*}[htb]
    \centering
    \includegraphics[width=0.9\textwidth]{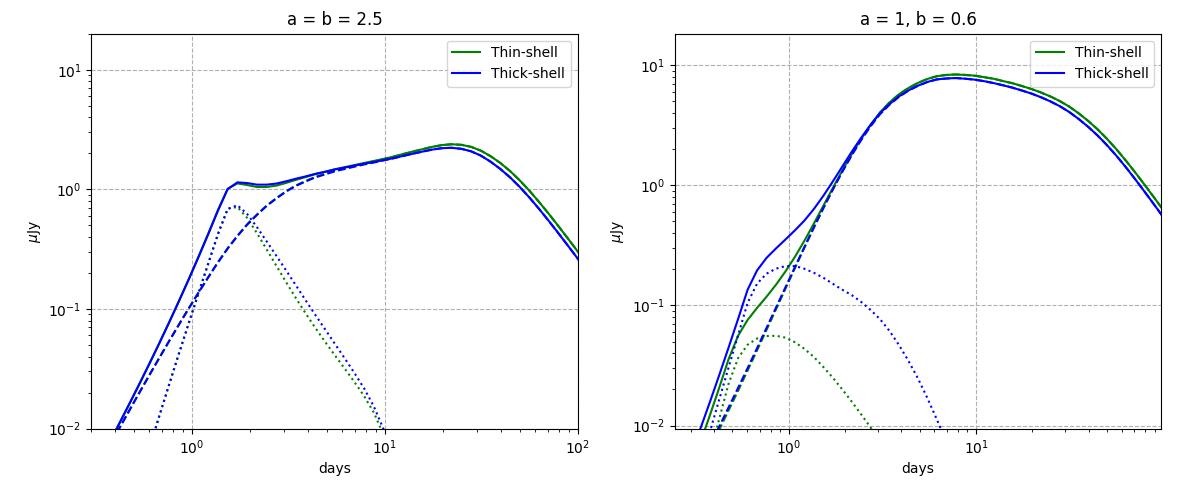}
    \caption{R-band afterglow for the thin-shell and thick shell case with different angular slopes.
    The dashed lines represent the FS emission, the dotted lines denote the RS emission, while the solid lines are total flux for each case. The width of thin-shell case is $\Delta_0 = 3\times 10^{10}~\mathrm{cm}$, whereas the width of the thick-shell case is $3\times 10^{12}~\mathrm{cm}$.}
    \label{Fig_thickness}
\end{figure*}

GRBs are commonly associated with relativistic jets, which can be approximated as shells at large radii. When discussing reverse shock (RS) emission, two scenarios are often considered: a thick shell and a thin shell.
However, for a misaligned structure jet, the light curve is more likely to resemble the thin shell case.
For instance, in a power-law jet model, the energy and initial bulk Lorentz factor distributions in the jet can be approximated as $\varepsilon(\theta)\propto\theta^{-a}$ and $\Gamma_0(\theta)\propto\theta^{-b}$.
The dimensionless quantity $\xi$ in Eq.~\ref{eq_xi}, which determines the shell thickness, varies with angle from the jet axis as $\xi(\theta)\propto \theta^{\frac{(4-k)b}{3-k}-\frac{a}{2(3-k)}}$. 
Under typical parameter values, $\xi(\theta)$ rapidly increases in the jet wing, resulting a thin shell that dominates the light curves at early times when viewed off-axis.
The comparison between light curves in the thin-shell and thick-shell case is illustrated in Fig.~\ref{Fig_thickness}. 
In the left panel, the thick shell case consists of a thick jet core surrounded by a thin jet wing, exhibiting no significant difference from the thin-shell case.
In the right panel, the condition $\xi(\theta)<1$ is satisfied throughout a jet for the thick-shell case, leading to more pronounced RS emission at early times.
Furthermore, the existing studies indicate that the thin shell model better describes on-axis observed afterglows compared to the thick shell model \citep{Japelj_2014,Gao_2015,Yi_2020}.

In this paper, we have applied the forward-reverse shock model to analyze the afterglow emission originating from an off-axis structured jet. 
We find that the RS emission may produce a prominent feature in misaligned structured jets at early times. 
In the mixed jet model viewed slightly off-axis, the RS emission may influence the optical and X-ray light curves during the decay phase and leads to a distinct feature before FS peak time in radio afterglow.
Utilizing this model, we have fitted the multi-wavelength afterglow light curves and VLBI proper motion of GW 170817/GRB170817A using four structured models: the two-component, power-law structured, Gaussian jet and mixed jet models. 
The fitting process is involved in the application of the MCMC method to obtain the best-fitting parameters.
Our analysis reveals that the early-time light curves are dominated by reverse shock emission and the gradual rise can be attributed to the jet structure and difference in the peak times of the reverse shock and forward shock in the baryonic jet.
In the case of a mixed jet, the reverse shock contributes only minimally to the light curve due to the relative low magnetic fields in reverse-shocked region. The initial gradual rise in this model is instead driven by the forward shock emission in the jet wing.
Notably, all four structured jet models yield consistent results for the viewing angle $\theta_\mathrm{obs}$ and jet half-opening angle $\theta_\mathrm{j}$, with values of approximately $\theta_\mathrm{obs}\approx17^\circ$ and $\theta_\mathrm{j}\approx6^\circ$. Additionally, the half-opening angle of the jet core is estimated to be $\theta_\mathrm{c}\lesssim5^\circ$.
Lastly, we find that the thick-shell case shows no significant difference from the thin-shell case under typical parameter values, though it can lead to brighter RS emission under specific conditions.

\begin{acknowledgments}
We would like to thank an anonymous referee for his/her helpful comments that have allowed us to improve our manuscript significantly. SLP thanks Prof. Rui-Zhi Yang for generously providing the computational resources on the server. These resources, offered without any form of compensation, have been crucial in advancing our research. This work was supported by the National SKA Program of China (grant No. 2020SKA0120300) and National Natural Science Foundation of China (grant No. 12393812).
\end{acknowledgments}


\vspace{5mm}


\software{emcee \citep{emcee}}


\clearpage
\appendix

\section{CORNER PLOTS OF THE FITTING RESULTS}
\begin{figure*}[hbt!]
    \centering
    \includegraphics[width=1.0\textwidth, angle=0]{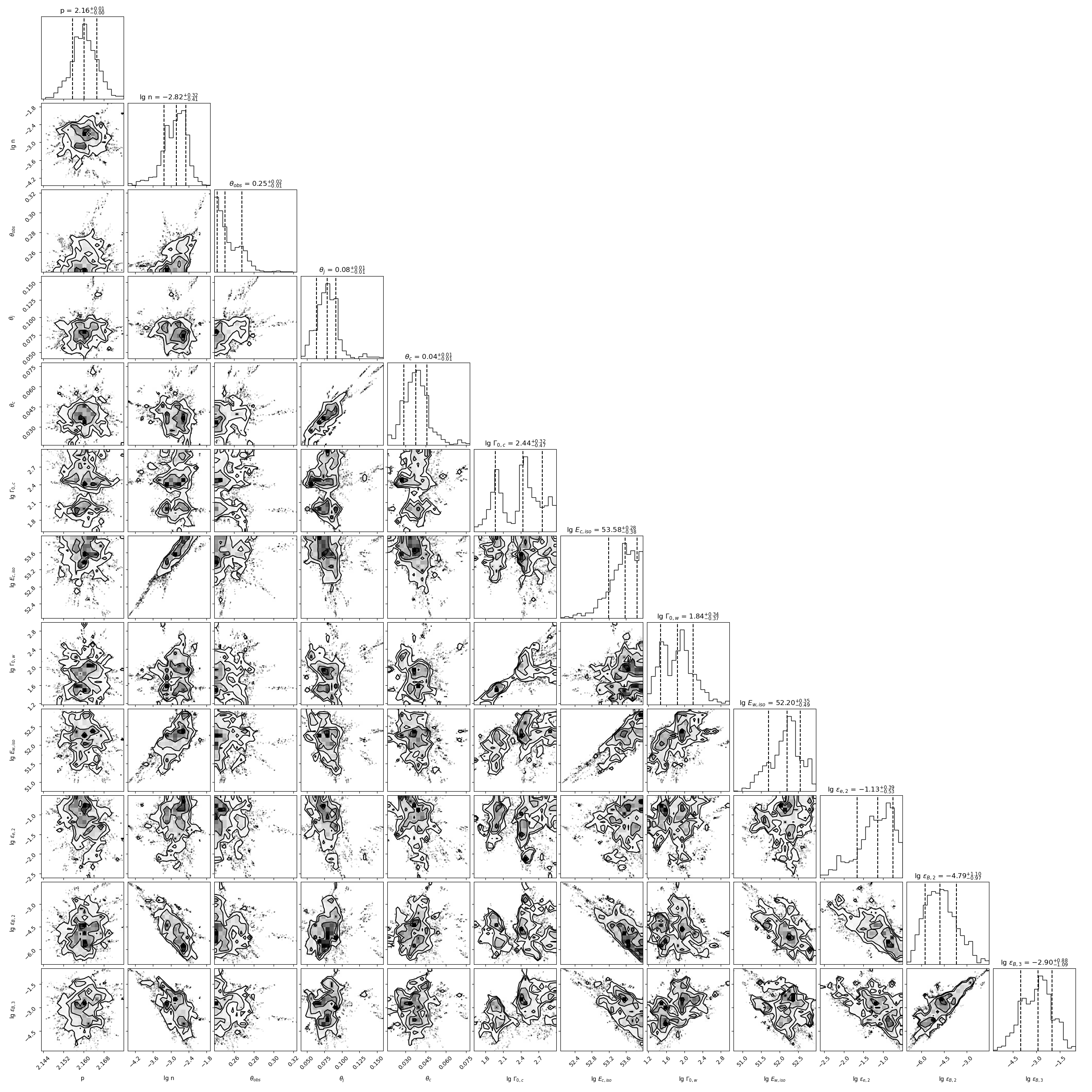}
    \caption{Corner plot of the parameters derived from fitting the multi-wavelength light curves of GRB 170817A with the two-component model. Our best-fitting parameters and corresponding 1$\sigma$ uncertainties are shown with the black dashed lines in the histograms on the diagonal.}
    \label{corner2C}
\end{figure*}

\begin{figure*}[h]
    \centering
    \includegraphics[width=1.0\textwidth, angle=0]{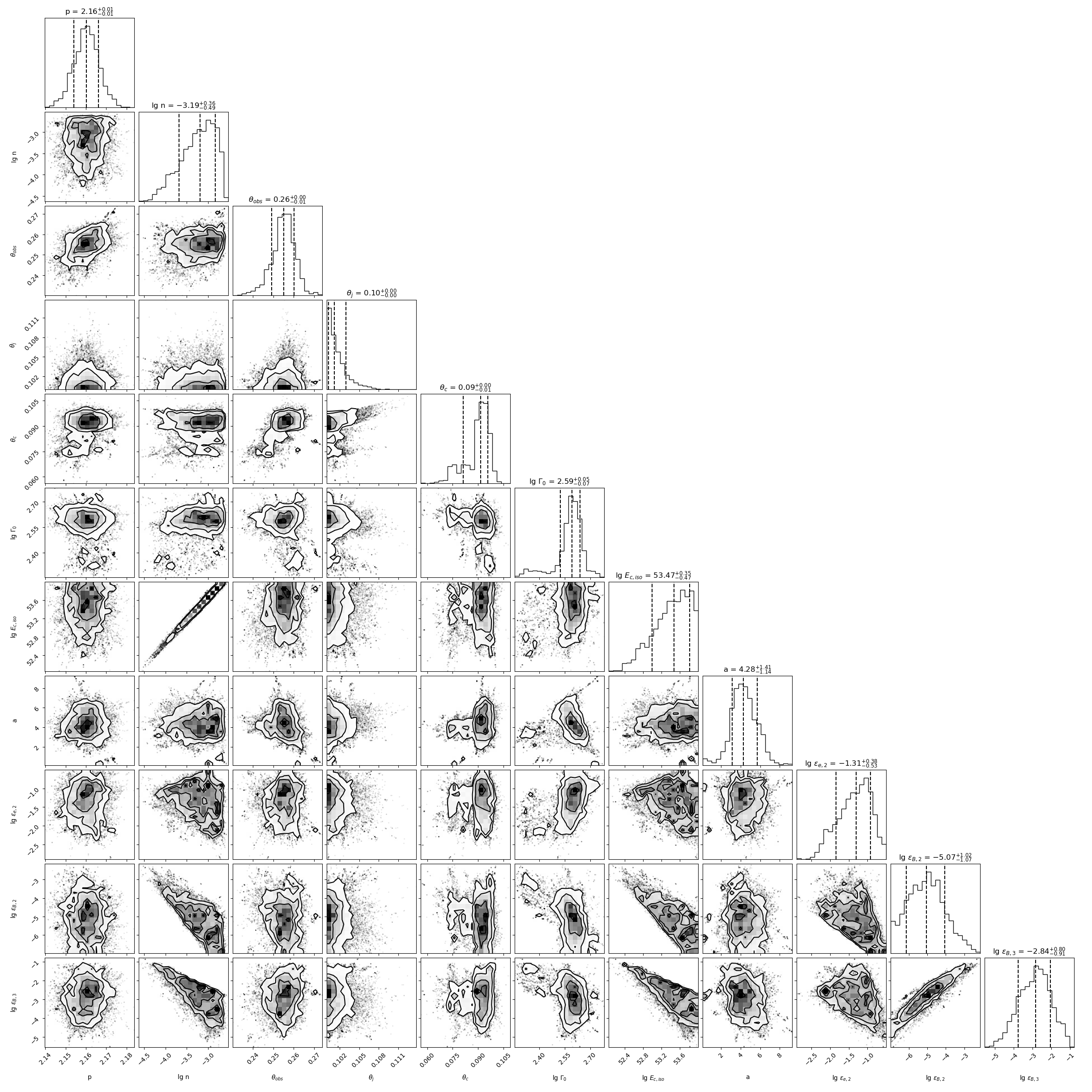}
    \caption{Same as Fig.~\ref{corner2C}, but for the power-law jet model.}
    \label{cornerPL}
\end{figure*}

\begin{figure*}[h]
    \centering
    \includegraphics[width=1.0\textwidth, angle=0]{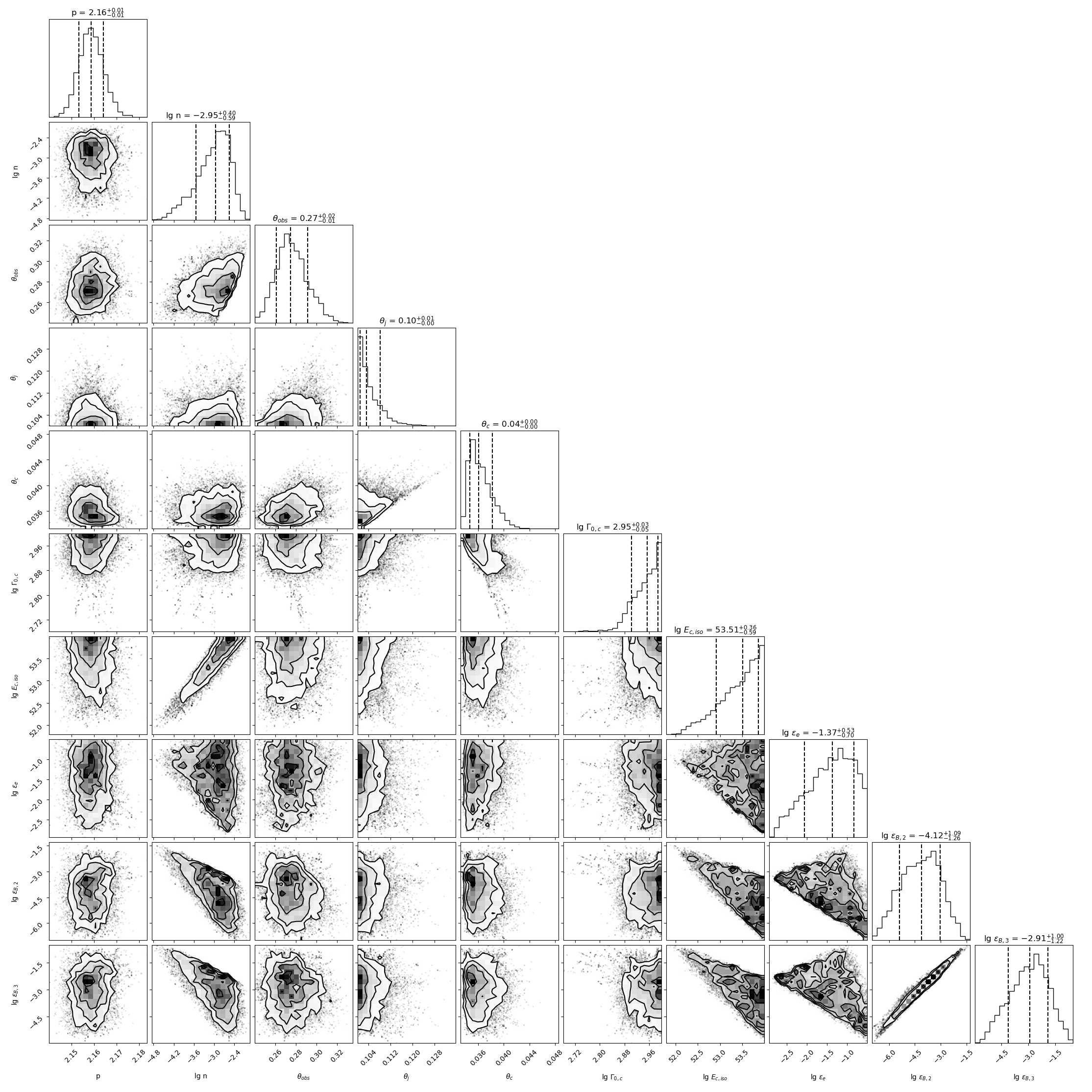}
    \caption{Same as Fig.~\ref{corner2C}, but for the Gaussian jet model.}
    \label{cornerG}
\end{figure*}

We also show the corner plots of the fitting results in Fig.~\ref{corner2C} - \ref{cornerG}.

\begin{figure*}[h]
    \centering
    \includegraphics[width=1.0\textwidth, angle=0]{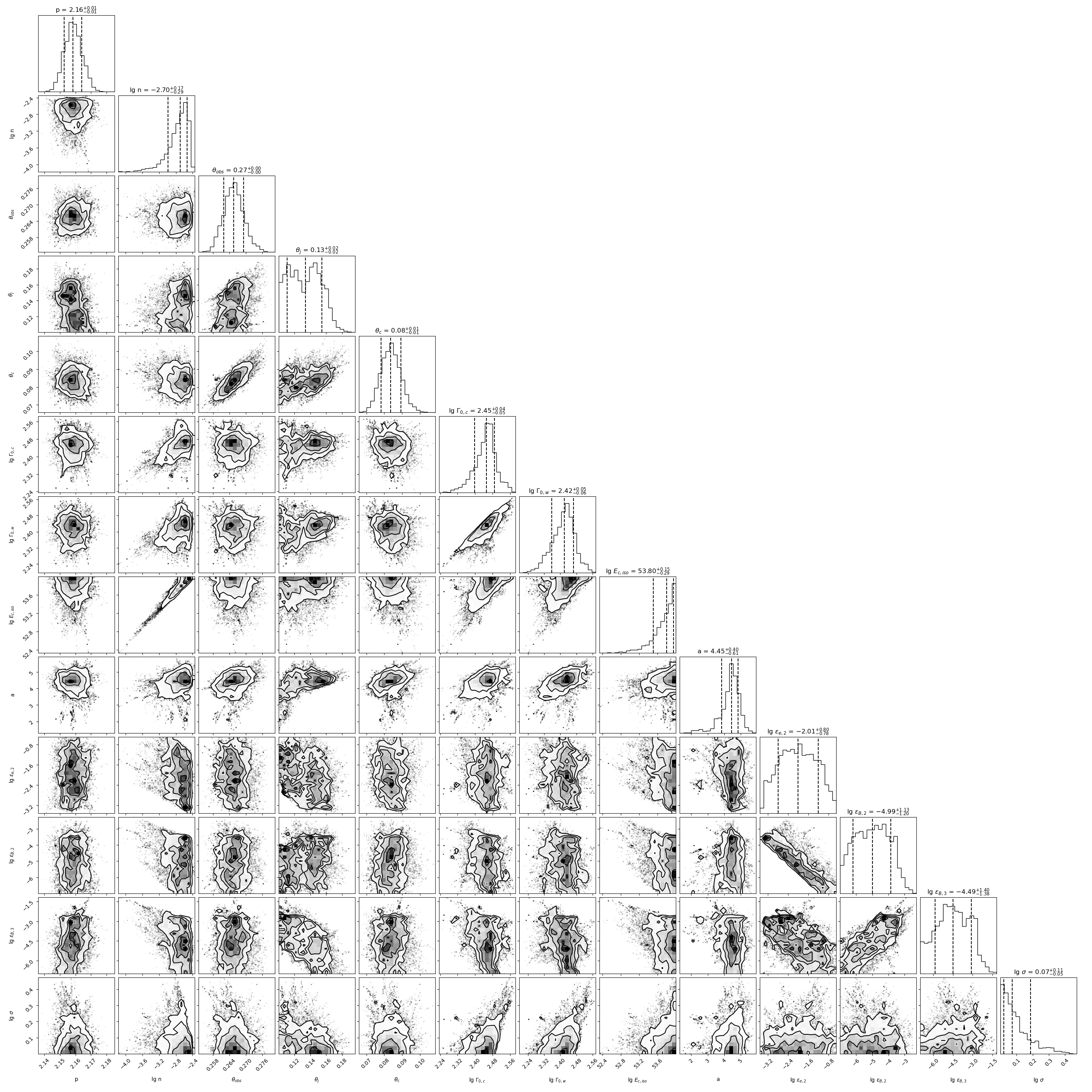}
    \caption{Same as Fig.~\ref{corner2C}, but for the mixed jet model.}
    \label{cornerMJ}
\end{figure*}

%Authors have the option to include names in Chinese, Japanese, or Korean (CJK) 
%characters in addition to the English name. The names will be displayed 
%in parentheses after the English name. The way to do this in AASTeX is to 
%use the CJK package available at \url{https://ctan.org/pkg/cjk?lang=en}.
%Further details on how to implement this and solutions for common problems,
%please go to \url{https://journals.aas.org/nonroman/}.

\clearpage

\bibliography{references}{}
\bibliographystyle{aasjournal}



\end{document}